\documentclass[lettersize,journal]{IEEEtran}
\usepackage[font=small,justification=centering]{caption}

\usepackage[caption=false,font=footnotesize,labelfont=sf,textfont=sf]{subfig}	
\usepackage{array}
\usepackage{graphicx}
\usepackage{algorithmic}
\usepackage{algorithm}
\usepackage{bm}
\usepackage{textcomp}
\usepackage{cite}
\usepackage{amsmath, amsfonts,amssymb, amsthm}
\usepackage{wasysym}
\usepackage{booktabs}
\usepackage{paralist}
\usepackage{balance}
\usepackage{multicol}
\usepackage{epsfig}
\usepackage{epstopdf}
\usepackage{upgreek}
\usepackage{color}
\usepackage{enumerate}
\usepackage{enumitem}
\usepackage{url}
\usepackage{verbatim}
\usepackage{graphicx}
\usepackage{mathrsfs}
\def\BibTeX{{\rm B\kern-.05em{\sc i\kern-.025em b}\kern-.08em
		T\kern-.1667em\lower.7ex\hbox{E}\kern-.125emX}}
\graphicspath{ {fig/} }

\setenumerate{leftmargin = 0em, itemindent = 15pt, listparindent = 1em}

\begin{document}

\title{Network-Aware Reliability Modeling and Optimization for Microservice Placement}
\author{Fangyu~Zhang,~\IEEEmembership{Graduate Student Member,~IEEE,},
		Yuang~Chen,~\IEEEmembership{Graduate Student Member,~IEEE,}
		Hancheng~Lu,~\IEEEmembership{Senior Member,~IEEE,},
		Yongsheng~Huang
\IEEEcompsocitemizethanks{
	\IEEEcompsocthanksitem Fangyu Zhang, Yuang Chen, Hancheng Lu, and Yongsheng Huang are with CAS Key Laboratory of Wireless-Optical Communications, University of Science and Technology of China, Hefei 230027, China (email: fv215b@mail.ustc.edu.cn; hclu@ustc.edu.cn; yuangchen21@mail.ustc.edu.cn; ysh6@mail.ustc.edu.cn).
}
}
\maketitle
\begin{abstract}
Optimizing microservice placement to enhance the reliability of services is crucial for improving the service level of microservice architecture-based mobile networks and Internet of Things (IoT) networks. Despite extensive research on service reliability, the impact of network load and routing on service reliability remains understudied, leading to suboptimal models and unsatisfactory performance. To address this issue, we propose a novel network-aware service reliability model that effectively captures the correlation between network state changes and reliability. Based on this model, we formulate the microservice placement problem as an integer nonlinear programming problem, aiming to maximize service reliability. Subsequently, a service reliability-aware placement (SRP) algorithm is proposed to solve the problem efficiently. To reduce bandwidth consumption, we further discuss the microservice placement problem with the shared backup path mechanism and propose a placement algorithm based on the SRP algorithm using shared path reliability calculation, known as the SRP-S algorithm. Extensive simulations demonstrate that the SRP algorithm reduces service failures by up to 29\% compared to the benchmark algorithms. By introducing the shared backup path mechanism, the SRP-S algorithm reduces bandwidth consumption by up to 62\% compared to the SRP algorithm with the fully protected path mechanism. It also reduces service failures by up to 21\% compared to the SRP algorithm with the shared backup mechanism.

\end{abstract}
\begin{IEEEkeywords}
	Microservice Placement, Reliability Model, Network State, Fault Tolerance, Shared Backup Path
\end{IEEEkeywords}

\IEEEdisplaynontitleabstractindextext

\IEEEpeerreviewmaketitle

\section{Introduction}
Cloud-native technologies \cite{usman2022survey} empower the creation and operation of applications in massively scalable distributed infrastructures, leveraging microservices architecture (MSA) \cite{soylemez2022challenges} alongside platform technologies \cite{kaur2022container} like containers and virtual machines. With the Cloud Native Computing Foundation's (CNCF) promotion of cloud-native technologies, MSA, which aims to improve software agility, is gradually coming into the limelight of both academia and industry. By splitting applications into microservices and interconnecting them using a lightweight application programming interface (API), MSA granulates complex services and provides an easier means of maintaining and updating software, accelerating new feature launches, and reducing manual costs. The benefits of MSA have led several major service providers such as Amazon, Netflix, and Spotify to use MSA to place their services \cite{soylemez2022challenges}, as well as making the Internet of Things (IoT) paradigm consider using MSA for smart manufacturing, Internet of Vehicles (IoV), and Industrial IoT (IIoT) \cite{siddiqui2023microservices}. Additionally, to fulfill service latency requirements \cite{kumar2023analysis} while avoiding the flooding of the backbone network with a large number of 5G and IoT devices \cite{chen2023statistical}, placing MSA-based services on infrastructure paradigms that are close to the users, such as edge or fog platforms \cite{kumar2023analysis}, is emerging as a new trend for service placement in 5G environments \cite{pallewatta2023placement}.

Ensuring the reliability of 5G application services is crucial for improving the users quality of experience (QoE), and the MSA complicates this issue. For ultra-reliable low-latency communication (URLCC) services in 5G, such as telemedicine and autonomous driving, end-to-end service reliability of five nines ($99.999\%$) or more is typically required \cite{zeng2023safedrl}. To improve reliability, traditional monolithic applications typically need to consider both the software reliability of the program itself and the reliability of the hardware on which the application is placed \cite{wang2021reliability}. However, with the introduction of MSA, placing microservices in a distributed manner means that the service needs to bear more risk of failure from both hardware and software \cite{liu2022reliability}. Therefore, how to improve the overall reliability of services when placing microservices has become an urgent problem.

A number of studies have been conducted to analyze the reliability model \cite{qiu2022online,zhu2022resource,rui2022multiservice,liu2022reliability,liu2022software,qiu2016hierarchical} and place microservices more reliably. Reliability models can be divided into two categories: hardware reliability models \cite{qiu2022online,zhu2022resource,rui2022multiservice} and software reliability models \cite{liu2022software,qiu2016hierarchical,schroeder2010large}. Based on the research on hardware and software reliability models, the reliability modeling studies for MSA-based services \cite{zhou2017cloud} comprehensively consider the overall reliability of the service after placing distributed software into the hardware. Since the placement strategy affects the service reliability, several works have been done to study microservice placement to enhance reliability \cite{zhu2023reliability, liu2021approach, huang2019reliable,ibrar2023reliability, martin2020crew,dadashi2023daip}. Microservice placement work can be categorized into placement in the cloud \cite{zhu2023reliability, liu2021approach, huang2019reliable} and placement in the edge or fog \cite{ibrar2023reliability, martin2020crew,dadashi2023daip,ramzanpoor2022multi} based on the application scenario. Due to the variety of application scenarios, they have addressed different issues in terms of resources, quality of service, and reliability, and hence differ in service reliability modeling.


However, the dynamic nature of the network state caused by network load \cite{yu2022online} and routing \cite{Markopoulou2008characterization} has not been well studied in service reliability modeling, which brings new challenges to microservice placement. Network load has been shown to be negatively correlated with hardware reliability, i.e., the higher the load, the lower the reliability. In this case, hardware reliability is always changing dynamically during microservice placement. Network routing refers to the routing between microservices. On the one hand, the hardware reliability on the communication path is also load-dependent. On the other hand, with the maturity of multipath routing technologies, multipath routing can also have a significant impact on service reliability \cite{le2023reliable, guima2022multipath, qu2020reliability}. As a result, the impact of dynamic network state caused by network load and routing cannot be ignored in service reliability modeling as well as microservice placement.

In this paper, we propose a network-aware service reliability model to address the aforementioned issue. Firstly, to consider the impact of network state changes on reliability modeling, the network state is sensed by building a load-dependent hardware node reliability model and a routing-dependent path reliability model. Then, as each microservice placement and routing between microservices may change the reliability of the infrastructure network, a network-aware placement algorithm is proposed to achieve optimal service reliability performance. Furthermore, to reduce the bandwidth consumption of the infrastructure network, we investigate the microservice placement problem with the shared backup path mechanism. In this case, shared backup path contention due to simultaneous backup failures brings the microservice placement problem a new network state change factor. Since contending paths may lead to routing failures of backup instances, the contention probability of shared backup paths is considered carefully when placing microservices. The main contributions of this article are summarized as follows:

\begin{itemize}
\item We propose a network-aware service reliability model to characterize the dynamic network states, with consideration of the load-dependent node reliability and the path reliability of multipath routing as well as the impacts of hardware and software decoupling and backup instances on the reliability of microservices. Simulation results validate the proposed reliability model with different algorithms in terms of the number of service failures.

\item Based on the proposed service reliability model, we formulate a microservice placement problem and then propose a service reliability-aware placement (SRP) algorithm to achieve maximum service reliability. The proposed algorithm evaluates the network-aware reliability of each microservice as it is placed. Simulation results show that the proposed algorithm reduces the number of service failures by up to 29\% compared to the benchmark algorithms.

\item To reduce bandwidth consumption, we further investigate the microservice placement problem with the shared backup path mechanism and propose an algorithm based on the SRP algorithm using shared path reliability calculation (i.e., SRP-S algorithm). The proposed algorithm approximates the contention probability by calculating the probability that a single failure causes contention on the shared backup path and then reduces the occurrence of contention by combining the probability with network-aware service reliability. Simulation results show that the SRP-S algorithm reduces the bandwidth consumption by up to 62\% compared to the SRP algorithm with the fully protected path mechanism and reduces the number of service failures by up to 21\% compared to the SRP algorithms with the shared backup path mechanism.
\end{itemize}
The rest of the paper is organized as follows. Section II discusses the work related to reliability modeling and reliability-aware microservice placement over protected or shared paths. Section III gives the modeling procedure for the system model and the proposed service reliability model. In Section IV, we formulate the microservice placement problem with the fully protected and shared backup paths. The corresponding algorithms to solve them are proposed in Section V and Section VI, respectively. Simulation results are discussed in Section VII. Finally, the paper concludes in Section VIII.
\vspace{-3mm}
\section{Related Work}
\vspace{-1mm}
Due to the distributed nature of MSA-based services, assessing service reliability is critical to ensure the failure tolerance of the MSA-based service, which has been addressed by research on reliability models and microservice placement.
\vspace{-5mm}
\subsection{Reliability Model}
\vspace{-1mm}
Studies on reliability models can be classified into two main categories based on failure causes: hardware reliability and software reliability \cite{qiu2016hierarchical}. For hardware reliability, most studies have modeled the arrival of hardware failures as Poisson processes \cite{tang2018queue}. Wang \textit{et al}. \cite{wang2021reliability} proposed an instance-sharing reliability model to aggregate multiple services into a composite and proposed an algorithm to improve its reliability. Zhu \textit{et al}. \cite{zhu2022resource} proposed a load-dependent node reliability model to capture the relationship between failure probabilities and workloads and introduced a recovery strategy to handle workload variations. Mtawa \textit{et al}. \cite{mtawa2021migrating} proposed a link reliability model to assess the all-pair reliability of the network and tested it on both conventional network and SDN. Similar to hardware failures, the arrival of software failures is also usually modeled as a Poisson process \cite{liu2022reliability}. Liu \textit{et al}. \cite{liu2022software} considered the software reliability problem in the framework of uncertainty theory and proposed a software reliability growth model based on uncertain differential equations.

Due to the different causes of hardware and software failures, it is inaccurate to consider the combination of hardware and placed software as a singular entity when modeling service reliability. Therefore, the service reliability model in the microservice placement problem should consider both hardware and software reliability \cite{qiu2016hierarchical, martin2020crew}. Qiu \textit{et al}. \cite{qiu2016hierarchical} investigated the reliability model and fault recovery of cloud computing platforms, where the reliability model considered both hardware reliability and software reliability. Martin \textit{et al}. \cite{martin2020crew} pointed out that the unavailability of a service is determined by software failures and hardware failures together and proposed a hardware-software decoupled reliability model. 
\vspace{-8mm}
\subsection{Microservice Placement}
\vspace{-1mm}
Microservice placement has been studied in a variety of scenarios \cite{liu2022reliability,zhao2022distributed,baranwal2022trappy}. Liu \textit{et al}. \cite{liu2022reliability} proposed an approach based on multi-intelligent body systems to maximize the reliability of services in cloud environments. Zhao \textit{et al}. \cite{zhao2022distributed} considered the heterogeneity of edge environments and the uncertainty of service requests. They modeled the microservice placement problem as a stochastic optimization problem and proposed a statistics-based approach to solve it. Baranwal \textit{et al}. \cite{baranwal2022trappy} investigated the truthfulness of fog owners in fog environments and proposed a heuristic algorithm to ensure the truthfulness of fog owners and the reliability of services.

Since the reliability of MSA-based services depends on the microservice placement strategy, there have been many studies that investigated placing microservices with the goal of improving service reliability \cite{zeng2023safedrl, martin2020crew, dadashi2023daip, ramzanpoor2022multi}. Zeng \textit{et al}. \cite{zeng2023safedrl} formulated the microservice placement problem as an integer nonlinear programming problem and proposed a deep reinforcement learning scheme based on expert intervention to ensure high reliability and low latency of the service. Martin \textit{et al}. \cite{martin2020crew} modeled the microservice placement problem as a multi-objective optimization problem and proposed a meta-heuristic algorithm to deal with the conflict between reliability and cost in the optimization objectives. Dadashi \textit{et al}. \cite{dadashi2023daip} enhanced the reliability of the service by using backups and proposed a reliability-aware and delay-efficient heuristic algorithm to solve the microservice placement problem. The authors in \cite{ramzanpoor2022multi} formulated the microservice placement problem in fog as a multi-objective optimization problem and proposed a fault-tolerant mechanism to improve the reliability of microservices while reducing power consumption and latency. However, while most works have used backup instances to enhance service reliability, the impact of network routing on service reliability has not been well studied.

Turning the perspective to routing, it can be seen that multipath routing techniques can significantly enhance the reliability of paths between microservices \cite{le2023reliable, guima2022multipath, qu2020reliability}. Le \textit{et al}. \cite{le2023reliable} proposed a reliable service provisioning scheme to optimize network resource utilization by using multipath routing. The authors in \cite{guima2022multipath} proposed a topology-agnostic multipath source routing scheme and orchestration architecture and verified its performance in improving communication reliability. Qu \textit{et al}. \cite{qu2020reliability} formalized the microservice placement problem as a mixed-integer linear programming problem and proposed a delay-aware hybrid multipath routing scheme to improve the reliability of network services.

To reduce bandwidth consumption, some researchers have considered using the shared backup path mechanism for network routing \cite{saidi2016resource, zheng2023robust, ergenc2021service}. Saidi \textit{et al}. \cite{saidi2016resource} proposed two shared path mechanisms, including shared backup paths and shared all paths, to conserve bandwidth resources during network routing. Zheng \textit{et al}. \cite{zheng2023robust} used shared backup path protection to improve bandwidth capacity limits for elastic optical networks and used backup paths to improve system reliability during network routing. Ergenc \textit{et al}. \cite{ergenc2021service} used shared backup paths for service placement. They asserted that the proposed shared backup capacity model can bring up to 70\% capacity gain and provide more than 90\% fault tolerance for single node failure.

In the aforementioned related work, few works are aware of the impact of the load state and routing state in the network on the service reliability model as well as the microservice placement strategy. In addition, the reliability gain of backup instances of microservices after the decoupling of hardware and software has also not been well studied.

\vspace{-3mm}
\begin{figure*}[!t]
	\centering
	\includegraphics[width=7in]{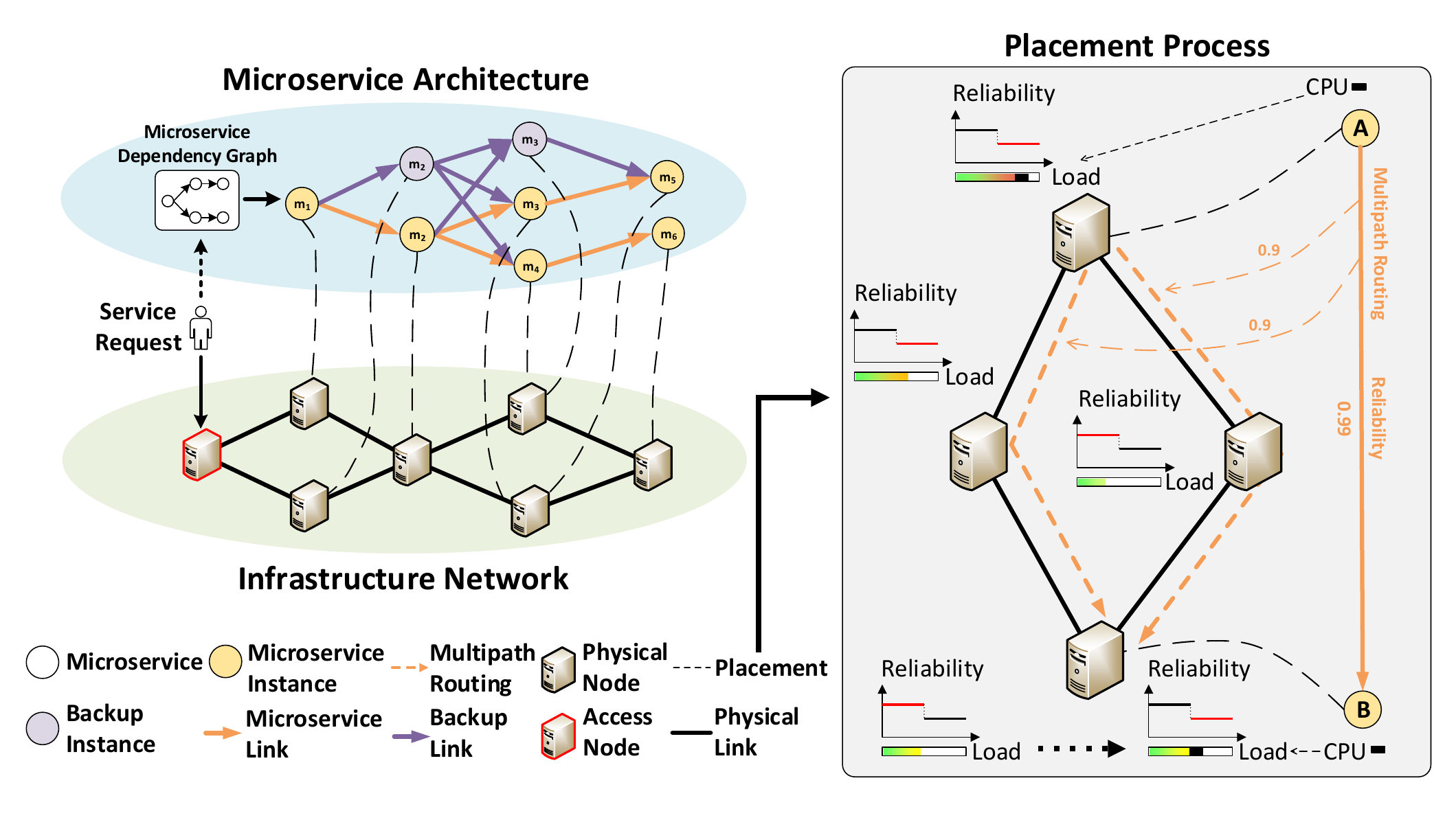}
	\caption{The process of placing microservices into the infrastructure network.}
	\label{fig0}
	\vspace{-5mm}
\end{figure*}
\section{System Model}
In this section, we give the system model and network-aware service reliability modeling. We first introduce the infrastructure network model for microservice placement and the service request model based on the MSA in Sec. \ref{net_model} and Sec. \ref{request_model}, respectively. Then, in order to clarify the difference between the already existing work and our work, we introduce the hardware reliability model and the software reliability model used in this paper in Sec. \ref{hard_and_soft_model}, which serves as the basis for our network-aware service reliability modeling. In Sec. \ref{service_model}, we formalize the network-aware service reliability model, which is innovatively sensitive to load-dependent node reliability and routing-dependent multipath routing reliability, and meticulously considers the impact of hardware and software reliability decoupling and backup instances on service reliability.
\vspace{-5mm}
\subsection{Infrastructure Network Model}\label{net_model}
\vspace{0mm}
The infrastructure network model is established to provide the underlying network for microservice placement. As shown in Fig. \ref{fig0}, we represent the infrastructure network with an undirected graph, $G = (N, E)$, where $N=\{n_1, n_2, \cdots, n_{|N|}\}$ denotes the set of physical nodes, and $E=\{e_{12}, e_{23}, \cdots, e_{(|N|-1)|N|}\}$ represents the set of physical links in the infrastructure network. For any physical node $n_i$, considering the most commonly utilized CPU core resources, we use $c(n_i)$ to denote the amount of resources that have been allocated, and $C(n_i)$ to represent its total resource capacity. For any physical link $e_{ij}$, the allocated bandwidth resources and total bandwidth resources are denoted by $bw(e_{ij})$ and $BW(e_{ij})$, respectively. Additionally, the reliability of physical nodes and links is denoted as $r_n$ and $r_e$, respectively, which describes the probability of no failure.
\vspace{-4mm}
\subsection{Service Request Model}\label{request_model}

The service request model is established to describe information related to service requests. We use $S$ to represent the set of all service requests. Each service request is represented as $s^i = (G^i, B^i, \Upsilon^i, D^i, \Omega^i)$, where $G^i$ is a directed acyclic graph (DAG) used to represent the microservices in the service request and their dependencies, $B^i$ denotes the maximum number of backups for microservices in the service request, $\Upsilon^i$ represents the amount of data transferred between microservices in service request $s^i$, $D^i$ denotes the set of latency deadlines for each microservice in the service request, and $\Omega^i$ denotes the lifetime of service request $s^i$. In the service request model described above, nodes in the microservice dependency graph $G^i$ represent microservices, and directed edges represent microservice links and invocation relationships between source and destination microservices. The microservice dependency graph $G^i$ and the microservice model with $B^i$ backup constraints are described in detail as follows.

The microservice dependency graph $G^i=(M^i, L^i)$ consists of the microservice set $M^i = \{m^i_0, m^i_1, m^i_2, \cdots, m^i_{|M|}\}$ and the microservice link set $L^i = \{l^i_{m_1m_2}, l^i_{m_2m_3}, \cdots, l^i_{m_{|M|-1}m_{|M|}}\}$. In microservice set $M^i$, $m^i_0$ is a special virtual microservice that represents the request access location and does not consume the computational resources of the access node. $m^i_1$ represents the root node of the microservice graph. Additionally, each microservice $m$ has fixed CPU core resource requirement $c(m)$ and reliability $r_m$. Microservice link $l$ has bandwidth resource requirement and reliability, represented as $bw(l)$ and $r_l$, respectively.

Microservice backups are placed as new instances of the primary microservices to enhance the overall reliability of the service. When the primary microservice instance is operating normally, backup microservice instances need to occupy computing resources on physical nodes and utilize bandwidth resources on physical links to provide failure tolerance. When the primary microservice instance fails, if a backup microservice instance is available, the microservice can still connect to upstream or downstream microservices through it, allowing the service to continue running. We use $m^i_j(b)$ to denote the $b$-th microservice instance of microservice $m^i_j$, where $b\in B^i_m$ and $B^i_m$ represents the set of backup instance indexes for microservice $m$. For convenience, we abbreviate the primary microservice $m^i_j(1)$ as $m^i_j$ in the following. In addition, $B^i$ in the service request model ensures that the number of microservice backups is limited to prevent unlimited resource consumption.

Since microservice instances typically have latency requirements, we consider that each microservice link has a latency, including transmission latency, propagation latency, and processing latency from child microservices, and represent the latency between a pair of parent-child microservices $d_{\tau m}$ as follows:
\begin{equation}\small
	d_{\tau m} = \frac{\upsilon^i_{\tau m}}{bw(l_{\tau m})} + \min_{p\in P_{l_{\tau m}}}\{\sum_{e\in p}d_e\}  + d_{m},
	\label{Lat}
\end{equation}
where $\tau$ is a parent microservice instance of $m$, $\upsilon^i_{\tau m}\in \Upsilon^i$ represents the amount of data transmitted from microservice $m$ to $\tau$ after processing, $d_e$ represents the propagation latency of physical link $e$, $P_{l_{\tau m}}$ denotes the set of paths where link $l_{\tau m}$ is placed, and $d_m$ represents the processing latency of microservice $m$. When a microservice link is placed on multiple paths, only the latency of the shortest path is considered.
\vspace{-3mm}
\subsection{Hardware And Software Reliability Model}\label{hard_and_soft_model}
\vspace{-1mm}
This subsection describes the reliability modeling of the elements involved in the microservice placement process, including hardware and software. First, the load dependency of hardware node reliability is the basis for service reliability to sense the network load, and the hardware link reliability constitutes the smallest unit of network routing-aware path reliability. Second, software reliability directly affects the service reliability gain of the backup instances. Additionally, software link reliability will directly affect the path reliability gained from multipath routing. The specific model is as follows:
\begin{enumerate}[label=$\mathbf{\arabic*}.$]
\item Hardware Reliability
	
The reliability of physical nodes has been demonstrated to be correlated with the workload running on them \cite{zhu2022resource, qiu2022online}. Load-dependent network node reliability affects the placement strategy, as densely placed microservices lead to lower reliability of physical nodes, while uniformly distributed microservices maintain a low load state of nodes at the cost of consuming more bandwidth resources. To describe the load dependence of dynamic network node reliability, we refer to the work of Zhu \textit{et al}. \cite{zhu2022resource} and represent the reliability $r_n$ of a physical node $n$ as a two-segmented function as follows:
	\begin{equation}\label{nodeR}\small
		r_n (c(n)) = \left\{
		\begin{aligned}
			r^L_n, \qquad&  c(n) \leq  \xi(n) \\
			r^H_n, \qquad&  \xi(n) < c(n) \leq C(n)
		\end{aligned}\qquad,
		\right.
	\end{equation}
where $\xi(n)$ represents the load threshold at which the reliability of the physical node changes, and $r^L_n$ and $r^H_n$ denote the reliability of the physical node under low-load and high-load conditions, respectively.
	
The failure arrival of a physical link $e$ is usually modeled as a Poisson process \cite{tang2018queue}. Therefore, we model its reliability $r_e$ as follows:
	\begin{equation}\small
		r_e(t) = Pr(x=0) = e^{-\lambda t}, 
	\end{equation}
where $Pr(\cdot)$ denotes probability, $\lambda$ represents the mean failure arrival rate, and $t$ denotes the time that the physical link has been operational.
	
\item Software Reliability
	
Software reliability encompasses the reliability of microservices and microservice links. Due to human factors that can lead to software failures, such as program design and environment configuration, software failures are typically modeled using specific software failure statistics. Nonetheless, the modeling of software failures does not affect subsequent analyses of service reliability, thus our work is compatible with arbitrary models. The arrival of software failures is modeled as a Poisson process. Therefore, the reliability $r_{m}$ and $r_{l}$ of microservice $m$ and microservice link $l$ can be respectively represented as
	\begin{equation}\small
		r_{m}(t) = Pr(x=0) = e^{-\lambda_1 t},
	\end{equation}
	\begin{equation}\small
		r_{l}(t) = Pr(x=0) = e^{-\lambda_2 t},
	\end{equation}
where $Pr(x=0)$ denotes the probability of no failure at moment $t$, and $\lambda_1$ and $\lambda_2$ are the mean failure arrival rates for microservices and microservice links, respectively.
\end{enumerate}
\vspace{-5mm}
\subsection{Network-Aware Service Reliability Model}\label{service_model}
In this subsection, network-aware service reliability is modeled to assess the reliability level of the microservice placement strategy. Network-aware service reliability considers network-aware reliability, reliability of microservice dependencies, and reliability gain of backup microservice instances. Among them, network-aware reliability specifically refers to routing-dependent multipath reliability consisting of load-dependent physical node reliability and physical link reliability.

First, we consider network-aware reliability. Since the reliability of a physical node is related to the load of microservices running on it, we define a binary variable $x^m_n$ to indicate whether microservice $m$ is placed on node $n$ or not, where $x^m_n=1$ means that microservice $m$ is placed on node $n$. In addition, we define an extra binary variable $y^l_e$ to indicate whether the microservice link $l$ is placed on the link $e$. Then we can use the defined binary variable $x^m_n$ to represent the physical node $n^m$ where microservice $m$ is placed as follows:
\begin{equation}\small
	n^m = \sum_{n\in N}x^m_nn.
\end{equation}
In addition, the CPU resource occupancy $c(n)$ can be represented as 
\begin{equation}\label{nodeU}\small
	c(n) = \sum_{s_i\in S}\sum_{m\in M^i}x^{m}_nc(m).
\end{equation}
Thus, the dynamic reliability of a physical node can be determined by Eq. (\ref{nodeR}) and Eq. (\ref{nodeU}). 

Path reliability is considered since not only does the operation of microservices require node reliability, but also microservice links require that all physical nodes and links in their paths are reliable. We denote the $j$-th path where the link between two microservice instances is placed by $p^{mm'}_j$, which is a set that contains all nodes and edges on the path, but not the source and destination nodes. We can then represent the path reliability $r_p$ for a single path $p$ as follows:
\begin{equation}\small
	r_p = \prod_{n \in N}r_n\prod_{e\in p}r_e.
\end{equation}
Next, we can represent the total path reliability $r_P$ of the path set which contains multiple paths $P$ as follows:
\begin{equation}\small
	r_P = 1-\prod_{p\in P}(1-r_p).
\end{equation}
For the subsequent calculations, we need to record the path information when calculating the reliability of each path or multiple paths. We define a function $\mathsf{P}(\cdot)$ that is used to query the corresponding path set from the calculated path reliability as follows:
\begin{equation}\small
	\mathsf{P}(r_P) = P,
\end{equation} 
where $P$ is the set of paths corresponding to the total path reliability $r_P$.

Next, we consider the total path reliability of a single microservice link placed on multiple paths. Calculating the two-terminal network reliability is the most accurate measure of total path reliability in a general network. However, in the microservice placement problem, routes are determined at the time of placement rather than being freely switchable at runtime to ensure resource provisioning. Therefore, we need to consider the reliability sum of a finite number of paths instead of two-end reliability. In this case, we consider all internally disjoint paths (IDPs) to avoid common cause faults (CCFs) \cite{rui2022multiservice}.

We propose a microservice path reliability matrix to describe the reliability of all IDPs between two microservices. First, we express the reliability of a physical node $n_i$ and the link $e_{ij}$ connected to it as $\mathsf{r}_{ij} = r_{n_i}r_{e_{ij}}$. Then, we propose a one-step path reliability matrix $R$ using the concept of the adjacency matrix as follows:
\begin{equation}\small
	R = R^{(1)} = 
	\begin{bmatrix}
		0 							     & \mathsf{r}_{12} 	 & \mathsf{r}_{13} & \cdots & \mathsf{r}_{1|N|}  & \\
		\mathsf{r}_{21}		& 0 							 & \mathsf{r}_{23} & \cdots & \mathsf{r}_{2|N|}  &\\
		\vdots						  & 							    &\ddots 				 &			  & \vdots					     & \\
		\mathsf{r}_{|N|1}	  & \mathsf{r}_{|N|2} & \cdots 				 &			   & 0								 & \\
	\end{bmatrix},
\end{equation}
where $\mathsf{r}_{ij} = 0$ if the physical link $e_{ij}$ does not exist. Then, for ease of representation in subsequent calculations, we define the path reliability operator as follows:
\begin{equation}\small
	\begin{aligned}
		x\!\oplus\! y &=1 - (1-x)(1-y) \\
		&= x+y -xy, \qquad x,y \in \left[0,1\right],
	\end{aligned}
\end{equation}
\vspace{-5pt}
\begin{equation}\small
	x\!\ominus\! y = \frac{x-y}{1-y}, \qquad x \in [0,1], y\in \left[0, x\right),
\end{equation}
\vspace{-5pt}
\begin{equation}\small
	\begin{aligned}
		&x\!\otimes \!y \!\!= \!\!\left\{\!
		\begin{aligned}
			&0, \quad p_x\!\cap\! p_y\neq \emptyset, p_y \!\in\!\mathsf{P}(y), \! p_x\!\in \!\mathsf{P}(x)  \\
			&xy, \quad\quad\quad\quad\quad else  
		\end{aligned},
		\right. 
		\\&\quad\quad\quad\quad\quad\quad\quad\quad\quad x, y\in [0, 1],
	\end{aligned}
\end{equation}
where $x\oplus y$ denotes the reliability sum of two paths, $x\ominus y$ denotes the reliability sum of multiple paths minus the reliability of one of the paths, and $x\otimes y$ denotes the reliability of two paths merged.

Based on the path reliability operator and the preservation of the path information corresponding to reliability, we define the multiplication of the path reliability matrix as follows:
\begin{equation}\small
	\begin{aligned}
		&\quad\quad\quad\quad\quad\quad\quad\quad  A \wedge B = C = [c_{ij}], \\
		&c_{ij} = \!\left\{
		\begin{aligned}
			&a_{i1}\otimes b_{1j} \oplus a_{i2} \otimes b_{2j} \cdots  a_{in}\otimes b_{nj}, \!\!\!\!\!& i\neq j\\
			&0, \!\!\!\!\!& i=j
		\end{aligned},
		\right.		
	\end{aligned}
\end{equation}
where $c_{ij}$ represents the reliability sum of equal-length paths from physical node $i$ to $j$. Thus, we can represent the $k$-th order path reliability matrix as follows:
\begin{equation}\small
	R^{(k)} = R^{(k-1)}\wedge R^{(1)},  k\geq 2,
\end{equation}
where $k$ represents the length of the path.

Next, we define path reliability matrix addition as follows:
\begin{equation}\small
	\begin{aligned}
		&\quad\quad\quad\quad\quad\quad\quad\quad A \vee B = C = [c_{ij}], \\
		&c_{ij}\!\! = \!\!\left\{
		\begin{aligned}
			&a_{ij}, \quad p_{a}\!\cap\! p_{b}\neq \emptyset, p_{a}\!\!\in\!\mathsf{P}(a_{ij}), p_{b}\!\!\in\! \mathsf{P}(b_{ij}) \\
			&a_{ij} \oplus b_{ij}, \quad \quad \quad \quad \quad else  
		\end{aligned}.
		\right.
	\end{aligned}
\end{equation}
As a result, we can represent all path reliability matrices with a maximum length of $k$ as follows:
\begin{equation}\small
	\hat{R}^{(k)} = R^{(1)}\vee R^{(2)}\vee \cdots\vee R^{(k)},
\end{equation}
where the elements $\hat{\mathsf{r}}^{(k)}_{ij}$ in $\hat{R}^{(k)}$ represent the reliability of all IDPs from physical node $n_i$ to physical node $n_j$ with a length not greater than $k$. However, the total path reliability is inaccurate because each path includes the source node within it and does not meet the definition of an IDP. Therefore, we denote the path set $\mathsf{P}'(\hat{\mathsf{r}}^{(k)}_{ij})$ after removing the source nodes of all paths as follows:
\begin{equation}\small
	\mathsf{P}'(\hat{\mathsf{r}}^{(k)}_{ij}) =\{p'|p'=p\backslash \{n_i\}, p \in \mathsf{P}(\hat{\mathsf{r}}^{(k)}_{ij})\}.
\end{equation}
Finally, we can represent the network-aware reliability matrix as follows:
\begin{equation}\small
	\mathcal{R}^{(k)} = \left[r^{(k)}_{ij}\right], r^{(k)}_{ij} = \left\{
	\begin{aligned}
		&r_{\mathsf{P}'(\hat{\mathsf{r}}^{(k)}_{ij})}, && i\neq j \\
		&1, && i=j
	\end{aligned}.
	\right.
\end{equation}
After obtaining the network-aware reliability matrix, we can analyze the reliability of the microservice dependency graph and the reliability gain brought by backup microservice instances. We divided the analysis of network-aware service reliability into the following four steps.

First, analyze the reliability between a single parent instance and a single child instance. We use $\tau_m^f$ to represent the $f$-th parent microservice of microservice $m$. For each child microservice instance, its parent microservice instance is connected to it through a microservice link placed on one or more paths. So we can consider the reliability of the microservice link from the $b_1$-th instance of the child microservice to the $b_2$-th instance of the $f$-th parent microservice and the network-aware reliability between the ends as a whole. We call this whole the effective probability of microservice link $\kappa^{(k)}_{m(b_1), \tau^f_m(b_2)}$ and denote it as follows:
\begin{equation}\small
	\kappa^{(k)}_{m(b_1), \tau^f_m(b_2)} =r_{l_{m(b_1)\tau^f_m(b_2)}}r^{(k)}_{n^{m(b_1)}n^{\tau^f_m(b_2)}},
\end{equation}
where $r_{l_{m(b_1)\tau^f_m(b_2)}}$ represent the software reliability of the microservice link and $r^{(k)}_{n^{m(b_1)}n^{\tau^f_m(b_2)}}$ represent the network-aware reliability between the nodes where the parent and child microservices are placed. In addition, since the virtual microservice $m_0$ used to represent the access location does not have a parent microservice, we let $\kappa_{m_0} = 1$.

Second, analyze the reliability of a single microservice instance and all its parent microservice links. To achieve this, understanding the impact of the node reliability on the effective probability of microservice links is essential. Node reliability may be reused by the network-aware reliability of multiple microservice instances and links, leading to spurious reliability. Therefore, node reliability should be considered only once in the calculation. However, backup instances introduce a new problem: node availability may be a non-essential condition for service availability. To address this issue, we focus on nodes whose failure would inevitably lead to service failure and call them critical nodes. We denote the set of critical nodes by $N^i_c$, which can be represented as follows:
\begin{equation}\small
	N^i_c \!  =\!\!   \{n'|n'=n,  \exists m \! \in \! M^i, n \! \! \prod_{b\in B^i_{m}}\! \! x^{m(b)}_n \!  \neq \! 0, n \! \in\!\!  N\},
\end{equation}
where $B^i_{m}$ denotes the set of backup instance indexes of microservice $m$. 

Now, we need to correct the network-aware reliability of the microservice link paths with the set of critical nodes. We denote the total path reliability $\hat{r}_P$ corrected by the set of critical nodes as follows:
\begin{equation}\small
	\hat{r}_P = \prod_{n \in p\backslash p\cap N^i_c}r_n\prod_{e\in p}r_e.
\end{equation}
Network-aware reliability can be corrected as follows:
\begin{equation}\small
	\hat{r}^{(k)}_{ij} = \left\{
	\begin{aligned}
		&\hat{r}_{\mathsf{P}'(\hat{\mathsf{r}}^{(k)}_{ij})}, && i\neq j \\
		&1, && i=j
	\end{aligned}.
	\right.
\end{equation}
The corrected effective probability of microservice link $\hat{\kappa}^{(k)}_{m(b_1), \tau^f_m(b_2)}$ can be expressed as follows:
\begin{equation}\small
	\hat{\kappa}^{(k)}_{m(b_1), \tau^f_m(b_2)} =r_{l_{m(b_1)\tau^f_m(b_2)}}\hat{r}^{(k)}_{n^{m(b_1)}n^{\tau^f_m(b_2)}}.
\end{equation}
Now we can denote the reliability of a single microservice instance $m(b)$ and all its parent microservice links by $\sigma^{(k)}_{m(b), n}$, which is as follows:
\begin{equation}\small
	\sigma^{(k)}_{m(b), n}\!\!=\!
	r_{\!m(b)}\!\!\!\!\!\! \prod_{f\in F_{m(b)}}\!\!\!\!(1 \!\!- \!\!\!\!\prod_{b' \in B_{\tau}}\!\! (1-  x^{m(b)}_{n} \hat{\kappa}^{(k)}_{m(b), \tau^f_{\!m}\!(b')})\!)
\end{equation}
where $B_{\tau}$ denotes the set of backup instance indexes of microservice $\tau^f_m$ and $F_{m(b)}$ denotes the set of indexes of the parent microservice instances. For convenience, we omit the superscript $(k)$ of $\sigma^{(k)}_{m(b), n}$ in the subsequent analyses, which simply denotes the maximum path length in the network-aware reliability. 

Third, analyze the reliability of all instances of a microservice and the links from their parents. We first use $\sigma_{m,n}$ to denote the reliability of all instances of microservice $m$ that are placed on the same node $n$, which can be expressed as follows:
\begin{equation}\small
	\sigma_{m,n} = 1-\prod_{b\in B_m}(1-\sigma_{m(b),n}).
\end{equation}
Now we can obtain the reliability of all instances of microservice $m$ on all nodes. We denote it by $\sigma_m$ as follows:
\begin{equation}\small
	\sigma_m = 1-\!\!\!\!\!\prod_{n\in N\backslash N^i_c}\!\!\!\!\!r_n(1-\sigma_{m,n}) \prod_{n\in N^i_c}(1-\sigma_{m,n}),
\end{equation}
where we temporarily disregard the reliability of critical nodes to avoid their reuse.

Fourth, analyze the reliability of the entire microservice dependency graph, i.e., network-aware service reliability. We denote the service reliability, which consists of the reliability of all microservices and the reliability of critical nodes, by $r_{G^i}$ as shown below:
\begin{equation}\small
	r_{G^i} = \prod_{n\in N^i_c}r_n\prod_{m\in M^i}\sigma_m.
\end{equation}

\vspace{-2mm}
\section{Problem Formulation}
\vspace{0mm}
The microservice placement problem is defined as a mapping $\psi: G_{i} \rightarrow G$. Microservice placement involves two tasks: node placement and path selection. Node placement involves placing each microservice instance (including backup microservice instances) from a service request onto a single physical node in the infrastructure network, while path selection involves mapping the link between any two microservice instances to one or more consecutive physical links. After a service request expires, microservice placement is revoked and the occupied resources are released. In this paper, the primary objective of microservice placement is to maximize the service reliability of a single service request while meeting latency and resource constraints. Therefore, for a single service request $s$ arriving at the current time, we can formalize the microservice placement problem as an integer nonlinear programming problem and represent it as follows:
\begin{equation}\small
		\mathcal{P}1: \max r_{G^i},
	\end{equation}
	\textit{s.t.}
	\begin{subequations}
		\begin{equation}\small
			\label{const_e}
			\sum_{s^i\in S}\sum_{l\in L^i}y^l_ebw(l) \leq BW(e), \forall e\in E,
		\end{equation}
		\begin{equation}\small
			\label{const_cpu}
			\sum_{s^i\in S}\sum_{m\in M^i}x^{m}_nc(m),\leq C(n), \forall n\in N,
		\end{equation}
		\begin{equation}\small
			\label{const_one}
			\sum_{n\in N} x^{m(b)}_n = 1, \forall  b\in B_m, m\in M,
		\end{equation}
		\begin{equation}\small
			\label{const_lat}
			d_{\tau m}\leq D_{\tau m}, \forall m\in M,
		\end{equation}
		\begin{equation}\small
			\label{const_backup}
			\sum_{m\in M}|B_m|\leq B + |M|,
		\end{equation}
		\begin{equation}\small
			\label{const_backup_num}
			B< |M|,
		\end{equation}
		\begin{equation}\small
			\label{xfield}
			x^m_n \in \left\{0, 1\right\}, \forall m\in M, n\in N,
		\end{equation}
		\begin{equation}\small
			y^l_e \in \left\{0, 1\right\}, \forall l\in L, e\in E,
		\end{equation}
		\begin{equation}\small
			c(n), c(m), C(n) \geq 0,  \forall m\in M, n\in N,
		\end{equation}
		\begin{equation}\small
			\label{bwfield}
			bw(l), bw(e), BW(e) \geq 0,  \forall l\in L, e\in E,
		\end{equation}	
	\end{subequations}
where constraints (\ref{const_e})-(\ref{const_cpu}) ensure that the resource requirements of service requests do not exceed the resource limits of physical nodes and links in the infrastructure network, constraint (\ref{const_one}) ensures that each microservice instance is placed on only one physical node, constraint (\ref{const_lat}) ensures that the latency of each microservice link does not exceed its latency requirements, constraint (\ref{const_backup}) ensures that the number of backup microservices does not exceed the backup limit of the service request, constraint (\ref{const_backup_num}) ensures that the backup limit does not exceed the number of microservices in the microservice dependency graph, and constraints (\ref{xfield})-(\ref{bwfield}) specify the value ranges of variables and resources.
	
In Section \ref{request_model}, we mentioned that backup microservice instances consume computing resources on physical nodes and bandwidth resources on physical links. However, since backup microservice instances mostly remain in an inactive state (becoming active only when the primary microservice instance fails), providing dedicated bandwidth protection for them is not always necessary. Therefore, we consider the concept of the shared backup path, which allows backup microservice instances to share network bandwidth resources to reduce bandwidth consumption. However, the introduction of the shared backup path mechanism creates a new problem: how to avoid multiple backup instances becoming active at the same time and causing network bandwidth contention, which can lead to service request failures. To solve this problem, we first introduce an upper limit, denoted as $\hat{BW}(e)$, which limits the shared bandwidth capacity. The upper limit is denoted as $\omega$ times the protected bandwidth limit $BW(e)$:
\begin{equation}\small
	\hat{BW}(e) = \omega BW(e),
\end{equation}
where $\omega \geq 0$. Then we modify constraint (\ref{const_e}) of problem $\mathcal{P}1$ and propose problem $\mathcal{P}2$, which aims to maximize service reliability with the shared backup path mechanism. Problem $\mathcal{P}2$ is as follows:
	\begin{equation}\small
	\mathcal{P}2: \max r_{G^i},
\end{equation}
\textit{s.t.}
\begin{subequations}
	\begin{equation}\small
		\label{const_shared_e}
		\sum_{s^i\in S}\sum_{l\in L^i_1}y^l_ebw(l) \leq BW(e), \forall e\in E,
	\end{equation}
	
	\begin{equation}\small
		\label{const_shared_backup_e}
\!\!\sum_{s^i\in S}\sum_{l\in L^i\!\backslash L^i_1}\!\!\!y^l_ebw(l) \!\leq\! \hat{BW}(e), \!\forall e\!\in\!\! E,
	\end{equation}
	
	\begin{equation}
		(\ref{const_cpu})-(\ref{bwfield}),
	\end{equation}
\end{subequations}
where $L^i_1$ represents the set of links between primary microservices. We denote it as follows:
\begin{equation}
	L^i_1 =  \{l|l=l_{m(1)m'(1)}, m, m'\in M^i\}.
\end{equation}
Constraint (\ref{const_shared_e}) ensures that protected bandwidth consumption does not exceed the protected bandwidth limit, while constraint (\ref{const_shared_backup_e}) guarantees that the shared bandwidth consumption does not exceed $\omega$ times the protected bandwidth limit.
\vspace{-4mm}
\section{Proposed SRP Algorithm}
\vspace{-1mm}
In this section, we propose a service reliability-aware placement (SRP) algorithm, which is a heuristic algorithm proposed to solve Problem $\mathcal{P}1$. The SRP algorithm takes as input the network state as well as the service request and outputs a microservice placement strategy that includes microservice instance placement and backup object selection.
\vspace{-5mm}
\subsection{Algorithm Description}
\vspace{-1mm}
The main process of the SRP algorithm is shown in Algorithm \ref{SRP}. When a service request arrives, SRP initiates a breadth-first search starting from the root microservice $m_1$ and adds the microservices in the microservice dependency graph $G^i$ to the placement queue $Q$ (line 3). In line 4, we define and initialize the backtracking counter $\epsilon$ with a predefined upper limit $\Delta$ and the node blacklist $N^m_{bl}$ for each microservice. The loop from line 5 to line 17 ensures that each microservice of the service request is placed. In line 6, the SRP extracts the first unplaced microservice $m$ from queue $Q$. The SRP then calls Algorithm \ref{MPAlg} to place the microservice $m$ and gets the placement result, which is used to indicate a successful or failed placement. Lines 8-16 deal with the case of microservice placement failure. If the microservice is not the root microservice and the number of backtracks has not exceeded the limit, SRP will cancel the placement of all parents of microservice $m$ and their children. It also adds the node where the parent was placed to the node blacklist of the parent and then proceeds to the next iteration. In line 18, we call Algorithm \ref{BuAlg} for backup object selection and backup instance placement. Finally, in line 19, the algorithm returns the placement success message.

Algorithm \ref{MPAlg} describes the microservice placement process. It traverses the nodes in the candidate node set (lines 4-22). In lines 5-7, if the node is in the blacklist of microservice $m$ or has insufficient resources, the algorithm starts the next iteration directly; otherwise, the algorithm calculates $\sigma_m$ for microservice $m$ in lines 8-16. Specifically, we use $n_{f_b}$ to denote the placement node of the parent microservice instance $\tau^f_m(b)$. In line 11, the algorithm searches for the set of IDPs between nodes $n_j$ and $n_{f_b}$ that meet the constraints. Line 12 calculates the total path reliability of the path set $P_{jf_b}$. Lines 13-16 calculate the reliability of $m$. Since node reliability is load-dependent, line 17 calculates the reliability of node $n$ after placing microservice $m$, and lines 18-22 consider the effect of the critical node set on service reliability. Lines 23-25 track the nodes with the highest total reliability. Line 27 places the microservice $m$ on the node with the highest total reliability and places all links connected to it on the path. At the same time, the algorithm records the current $\sigma_m$ for subsequent selection of backup objects. Since the computation of the current $\sigma_m$ is performed simultaneously with the placement of the microservice links, no additional time complexity is added. Finally, lines 28-32 return the placement result.

Algorithm \ref{BuAlg} outlines the strategy for selecting backup objects. After initializing the backup counter $b$ and the set of backup objects $BM^i$ in line 1, the algorithm enters a loop in lines 2-16, which requires that the number of backup instances does not exceed a limit and that the set of backup objects is not empty. In lines 4-8, the algorithm iterates over $BM^i$ to obtain the microservice $m_{min}$ with the smallest $\sigma_m$ value. Since the instantaneous $\sigma_m$ obtained when placing the microservice increases with the number of instances, the latest $\sigma_m$ needs to be obtained in line 5. The algorithm then calls Algorithm \ref{MPAlg} in line 10 to place the microservice $m_{min}$. In lines 11 to 15, if the placement fails, the microservice is removed from the set of backup objects and the next round of iteration starts; otherwise, the backup counter is increased in line 15.
 \begin{algorithm}[!t]
	\caption{Service Reliability-aware Placement (SRP) Algorithm}
	\label{SRP}
	\begin{algorithmic}[1]
		\STATE \textbf{Input} edge network $G$, service request $s^i$.
		\STATE \textbf{Output} placement result of $G^i$ with backups.
		
		\STATE Add microservices from $M^i\backslash{m^i_0}$ to the placement queue $Q$ starting from $m_1$ through breadth-first search.
		\STATE $\epsilon \leftarrow  0$, $N^m_{bl} \leftarrow \emptyset, m\in Q$.
		\WHILE {unplaced microservices in $Q$ exist}
		\STATE Obtain the first unplaced microservice $m$ from $Q$.
		\STATE $res\leftarrow$ place $m$ using Alg. \ref{MPAlg}.
		\IF{$res =\FALSE$}
		\IF{$m \neq m_1$ \AND $\epsilon < \Delta$}
		\STATE Undo the placement of all the parents of microservice $m$ and all their children. 
		\STATE $\epsilon\leftarrow \epsilon + 1, N^{\tau_m}_{bl} \leftarrow N^{\tau_m}_{bl}\cup \{n^{\tau_m}\}$ 
		\STATE \textbf{continue}
		\ELSE
		\RETURN \FALSE
		\ENDIF
		\ENDIF
		\ENDWHILE
		\STATE Backup Placement Process (Alg. \ref{BuAlg}).
		\STATE \textbf{return}  \TRUE
	\end{algorithmic}
\end{algorithm}

\begin{algorithm}[!t]
	\caption{Microservice Placement Process}
	\label{MPAlg}
	\begin{algorithmic}[1]
		\STATE \textbf{Input} $G$, $s^i$, $m$, $N^m_{bl}$.
		\STATE \textbf{Output} Placement result of microservice $m$.
		\STATE $n_{max} \leftarrow null, \mathsf{r}_{max} \leftarrow 0.$
		\FOR{$n_j\in N$}
		\IF{$n_j\in N^m_{bl}$ \textbf{or} $n_j$ is under-resourced}
		\STATE \textbf{continue}
		\ENDIF
		\STATE $\mathsf{r}_{b} \leftarrow0, \mathsf{r}_{f} \leftarrow1$.
		
		\FOR{$f \in F_m$}
		\FOR{$b\in B_{\tau^f_m}$}
		\STATE Denote the node of $\tau^f_m(b)$ as $n_{f_b}$ and get the set of paths $P_{jf_b}$ that meet the constraints.
		\STATE Calculate the total path reliability $r_{P_{jf_b}}$.
		\STATE $\mathsf{r}_{b}\leftarrow  \mathsf{r}_{b} \oplus r_{P_{jf_b}}$.
		\ENDFOR
		\STATE $\mathsf{r}_{f}\leftarrow\mathsf{r}_{f} \mathsf{r}_{b}$.
		\ENDFOR
		
		\STATE Calculate the $r'_{n_j}$ after placing $m$ to $n_j$.
		\IF{$n_j \in N_c$}
		\STATE $\mathsf{r}_{f}\leftarrow\mathsf{r}_{f} \frac{r'_{n_j}}{r_{n_j}}$.
		\ELSE
		\STATE $\mathsf{r}_{f}\leftarrow\mathsf{r}_{f}r'_{n_j}$.
		\ENDIF
		\IF{$\mathsf{r}_{max} < \mathsf{r}_{f}$}
		\STATE $\mathsf{r}_{max} \leftarrow\mathsf{r}_{f}, n_{max} \leftarrow n_j$.
		\ENDIF
		\ENDFOR
		\STATE Place $m$ to $n_{max}$ while recording the current $\sigma_m$.
		\IF{placement succeeds}
		\STATE \textbf{return} \TRUE
		\ELSE
		\STATE \textbf{return} \FALSE
		\ENDIF
	\end{algorithmic}
\end{algorithm}
\vspace{7mm}
\begin{algorithm}[!t]
	\caption{Backup Placement Process}
	\label{BuAlg}
	\begin{algorithmic}[1]
		\STATE $b\leftarrow 0, BM^i \leftarrow M^i\backslash\{m^i_0\}$.
		\WHILE{$b<B^i $ \textbf{and} $ BM^i \neq\emptyset$ }
		\STATE $m_{min} \leftarrow null, r_{min} \leftarrow1$.
		\FOR{$m\in BM^i$}
		\STATE Obtain the $\sigma_m$ of the last instance of $m$.
		\IF{$r_{min} > \sigma_m$}
		\STATE $r_{min} \leftarrow \sigma_m, m_{min}\leftarrow m$.
		\ENDIF
		\ENDFOR
		\STATE $res\leftarrow$ place $m_{min}$ using Alg. \ref{MPAlg}.
		\IF{$res=\FALSE$}
		\STATE $BM^i \leftarrow BM^i\backslash\{m_{min}\}$.
		\STATE \textbf{continue}
		\ENDIF
		\STATE $b\leftarrow b+1$.
		\ENDWHILE
	\end{algorithmic}
\end{algorithm}

\vspace{-11mm}
\subsection{Complexity Analysis}
The complexity analysis starts with the time complexity of Algorithm \ref{MPAlg} and Algorithm \ref{BuAlg}. First, we use the IDP-searching algorithm based on the Edmonds-Karp algorithm \cite{edmonds1972theoretical} with a time complexity of $O(|N||E|^2)$ to obtain the set of paths between two nodes. Then we can get the complexity of Algorithm \ref{MPAlg}, which is $O(|N|^2 |E|^2 |M|)$. In Algorithm \ref{BuAlg}, the loop in lines 2 to 16 is executed at most $|M|$ times. Since the complexity of each loop is $O(|M|+|N|^2 |E|^2 |M|)=O(|N|^2 |E|^2|M|)$, the total complexity of Algorithm \ref{BuAlg} is $O(|N|^2|E|^2|M|^2)$. Now looking back at Algorithm \ref{SRP}, we can see that its complexity is the same as that of Algorithm \ref{BuAlg}. Therefore, we can conclude that the complexity of Algorithm \ref{SRP} is $O(|N|^2|E|^2|M|^2)$.

\vspace{-3mm}
\section{Proposed SPRC Algorithm}
\vspace{-1mm}
In this section, we propose a new heuristic algorithm based on the SRP algorithm to solve the problem $\mathcal{P}2$ by using the shared path reliability computation (SPRC) algorithm. When selecting placement nodes for backup instances, in addition to considering the adequacy of the remaining shared bandwidth on the physical link, we also need to consider the shared backup path contention problem caused by simultaneous failures. Specifically, when other backup links are suddenly activated (i.e., switched from occupying virtual bandwidth to occupying protected bandwidth), the protected bandwidth may be insufficient and lead to the failure of the backup link if the current backup link needs to be activated for failure tolerance. Therefore, we must consider the reliability of the primary instance of the backup link placed on the physical link, which determines the distribution of the occupied bandwidth of the protected bandwidth. For a more comprehensive assessment of network-aware service reliability, we propose replacing lines 9-16 in Algorithm \ref{MPAlg} with SPRC. In the following, we refer to the SRP algorithm using SPRC as the SRP-S algorithm.

\vspace{-4.3mm}
\subsection{Algorithm Description}
\vspace{-1mm}

Algorithm 4 modifies lines 9-16 of Algorithm 2. In lines 6-11 of Algorithm 4, we traverse the set $Mp$, which represents the set of all backup microservice instances that have links on the path $p$. Line 7 obtains the probability of inactivity of the backup microservice instance $m'$, which is the reliability of all instances with backup indexes less than the backup index of $m'$. In lines 8-10, we evaluate whether the protected bandwidth is sufficient on the physical link for the microservice links belonging to both $m'$ and $m$. Activating $m'$ may cause activation of $m$ to fail if bandwidth is insufficient. In fact, each backup instance in $M'$ may be inactive or active, so there are $2^{|M'|}$ events and there is only one event with the highest probability that each backup instance is inactive. Due to the time constraint and the low probability of multiple simultaneous failures, we ignore the simultaneous failure of two or more backup instances in this algorithm and only consider $|M'|$ individual failure events. Finally, in line 12, we multiply the backup path inactivation probability by the original path reliability to produce new path reliability and calculate $\hat{r}_{P_{jf_b}}$ with the new path reliability in line 14. Subsequent lines 15-17 are consistent with Algorithm \ref{MPAlg}.
\vspace{-5mm}
\subsection{Complexity Analysis}
\vspace{-1mm}
The additional complexity introduced by Algorithm 4 relative to Algorithm 2 is mainly in the loops in lines 6-11. Combined with the complexity of the path-searching algorithm, we can obtain the time complexity of Algorithm 4 as $O(|M|(|N||E|^2+|M|))$. Thus the time complexity of the SRP-S algorithm is $O(|M|(|M| + |N||M|(|N||E|^2 + |M|))) = O(|N|^2|E|^2|M|^2+|N||M|^3)$.

 \begin{algorithm}[!t]
	\caption{Shared Path Reliability Calculation (SPRC) Algorithm}
	\label{SPRCAlg}
	\begin{algorithmic}[1]
		\FOR{$f \in F_m$}
		\FOR{$b\in B_{\tau^f_m}$}
		\STATE Denote the node of $\tau^f_m(b)$ as $n_{f_b}$ and get the set of paths $P_{jf_b}$ that meet the constraints.
		\FOR{$p\in P_{jf_b}$}
		\STATE $Pr_{p} \leftarrow 0$.
		\FOR{$m'\in M_{p}$}		
		\STATE Obtain the inactivation probability $\sigma'_{m'}$ for $m'$.
		\IF{$\exists e\in p\cap\{e|y^{l_{m'\tau}}_e = 1\}, BW(e) < bw(e) + bw(l_{m'\tau}) + bw(l_{m\tau^f_m(b)})$}
		\STATE $Pr_{p} \leftarrow Pr_{p} +\sigma'_{m'}$.
		\ENDIF
		\ENDFOR
		\STATE $\hat{r}_p \leftarrow Pr_p r_p$.
		\ENDFOR
		\STATE Calculate $\hat{r}_{P_{jf_b}}$ using modified path reliability.
		\STATE $\mathsf{r}_{b}\leftarrow  \mathsf{r}_{b} \oplus \hat{r}_{P_{jf_b}} $.
		\ENDFOR
		\STATE $\mathsf{r}_{f}\leftarrow\mathsf{r}_{f} \mathsf{r}_{b}$.
		\ENDFOR
	\end{algorithmic}
\end{algorithm}
\vspace{-2mm}
\section{Performance Evaluation}
We validate our modeling and algorithmic work through extensive simulations. Our simulation code can be accessed online \cite{zhang2024reliability}.
\vspace{-4mm}
\subsection{Simulation Setting}
\vspace{-1mm}
We first use the Erdős-Rényi model \cite{erdds1959random} to create an infrastructure network topology with a node count of 50 and an edge creation probability of 0.2. We then select one-fifth of the nodes with smaller degrees as access nodes for receiving service requests. For each microservice, microservice link, and physical link, we set their failure arrival rate to $0.00001$ per time unit, ensuring their reliability meets the "five nines" reliability level at the initial moment. For dynamic node reliability, we set the reliability $r_L$ between $0.9999$ and $0.99999$ for low load and $r_H$ between $0.999$ and $0.9999$ for high load. For the backup number limit, we set it to be the same as the primary number of microservices (full backup) if not specifically declared. The other parameters are listed in Table \ref{paras}. In all simulations, $100$ service requests reach the access node according to a Poisson distribution with parameter $1$, implying that the average arrival rate of service requests is $1$ per time unit. Additionally, all simulations are performed $100$ times and averaged for the final results.
\begin{table}[!t]
	\renewcommand{\arraystretch}{1.3}
	\caption{Parameters}
	\label{paras}
	\centering
	\begin{tabular}{|c|c|c|}
		\hline
		\textbf{Element} & \textbf{Parameter} & \textbf{Range} \\
		\hline	$n$& $C(n)$ & $[8,16]$ cores \\
		\cline{2-3}
		& $\xi(n)$ & $0.5$ \\
		\hline
		& $BW(e)$ & $[100, 1000] $ MBps\\
		\cline{2-3}
		$e$
		& $d_e$ & $[1, 10]$ ms\\
		\cline{2-3}
		& $\omega$ & $1$ \\
		\hline
		
		$s$& $\Omega$ & $[1, 100]$  units\\
		\cline{2-3}
		 & $|M|$ &$[1,5]$ \\
		\hline
		& $c(m)$ & $[0.1,1]$  cores\\
		\cline{2-3}
		$m$ & $\upsilon_{\tau m}$ &$[0.1,5]$ MB \\
		\cline{2-3}
		& $d_m$ & $[10, 50]$ ms \\
		\hline
		$l$& $bw(l)$ & $[0.1,10]$ MBps \\
		\cline{2-3}
		& $D_l$ & $[0.03, 50.15]$ s \\
		\hline
	\end{tabular}
\end{table}
\vspace{-5mm}
\subsection{Benchmark}
In our simulations, four algorithms are used for comparison with the SRP and SRP-S algorithms. They are described in detail as follows:
\begin{enumerate}[label=$\mathbf{\arabic*}.$]
	\item Delay-efficient and Availability-aware Placement (DAIP) \cite{dadashi2023daip}

	DAIP algorithm is an algorithm that considers backup instances. It focuses on the reliability of the nodes when placing services and chooses the path with the least latency when placing links. In addition, it adopts a Round-Robin strategy for backup object selection. While the DAIP algorithm considers in detail the reliability gain from backup instances, it does not consider in detail the dynamic network-aware reliability and hardware-software reliability decoupling in contrast to our work.
	
	\item Reliable Redundant Services Placement (RRSP) \cite{huang2019reliable}
	
	The RRSP algorithm considers only the total node reliability after placement and does not consider path selection. Unlike DAIP, RRSP does not focus on the placement of each instance but directly generates a certain number of solutions and selects the optimal solution among them. In addition, it does not specify how the backup objects are selected, so we generate backup instances for the primary instances in descending order of the degrees of the nodes in the dependency graph. We use this benchmark algorithm to show the performance of the algorithm considering only node reliability.
	
	\item Greedy Placement (Grd)
	
	The greedy algorithm is a classical heuristic algorithm. It selects the node with the highest reliability for each microservice instance and chooses the shortest path for each microservice link. To emphasize the effect of backups on reliability improvement, the basic version of the greedy algorithm does not consider the backup algorithm.
	
	\item Greedy Placement with Backup (Grd-B)
	
	Greedy Placement with Backup is an advanced version of Greedy Placement where the round-robin policy is used for backup object selection. The instance placement and path selection of this algorithm are consistent with Greedy Placement.
\end{enumerate}
\vspace{-5mm}
\subsection{Validation of Network-Aware Service Reliability Model}\label{sim_model}
\vspace{-1mm}
\begin{figure}[!t]
	\centering
	\subfloat[Service request distribution for different service reliability.]
	{\includegraphics[width=1.5in]{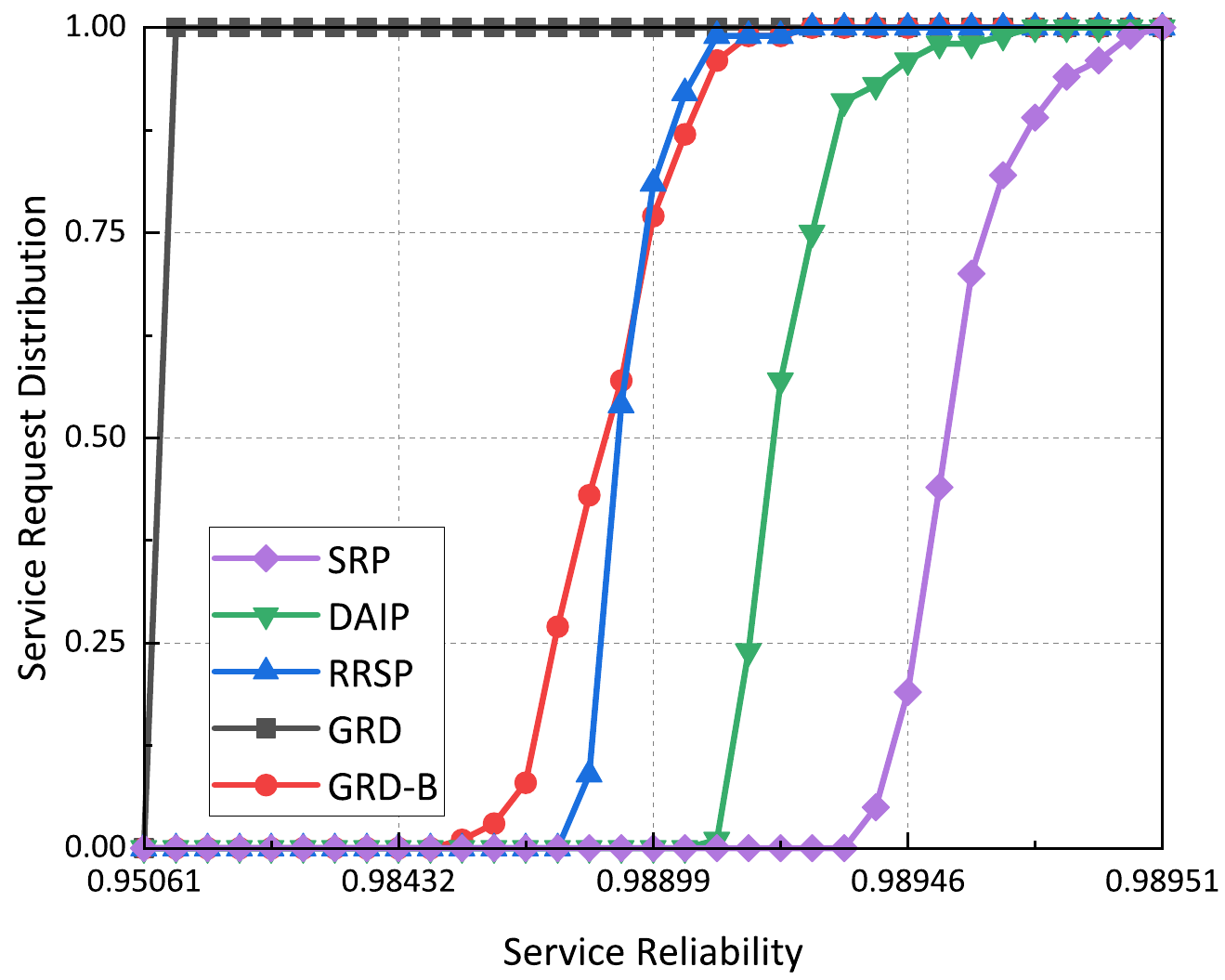}
		\label{fig1a}}
	\hfil
	\subfloat[Number of service failures over time.]
	{\hspace{-3mm}\includegraphics[width=1.48in]{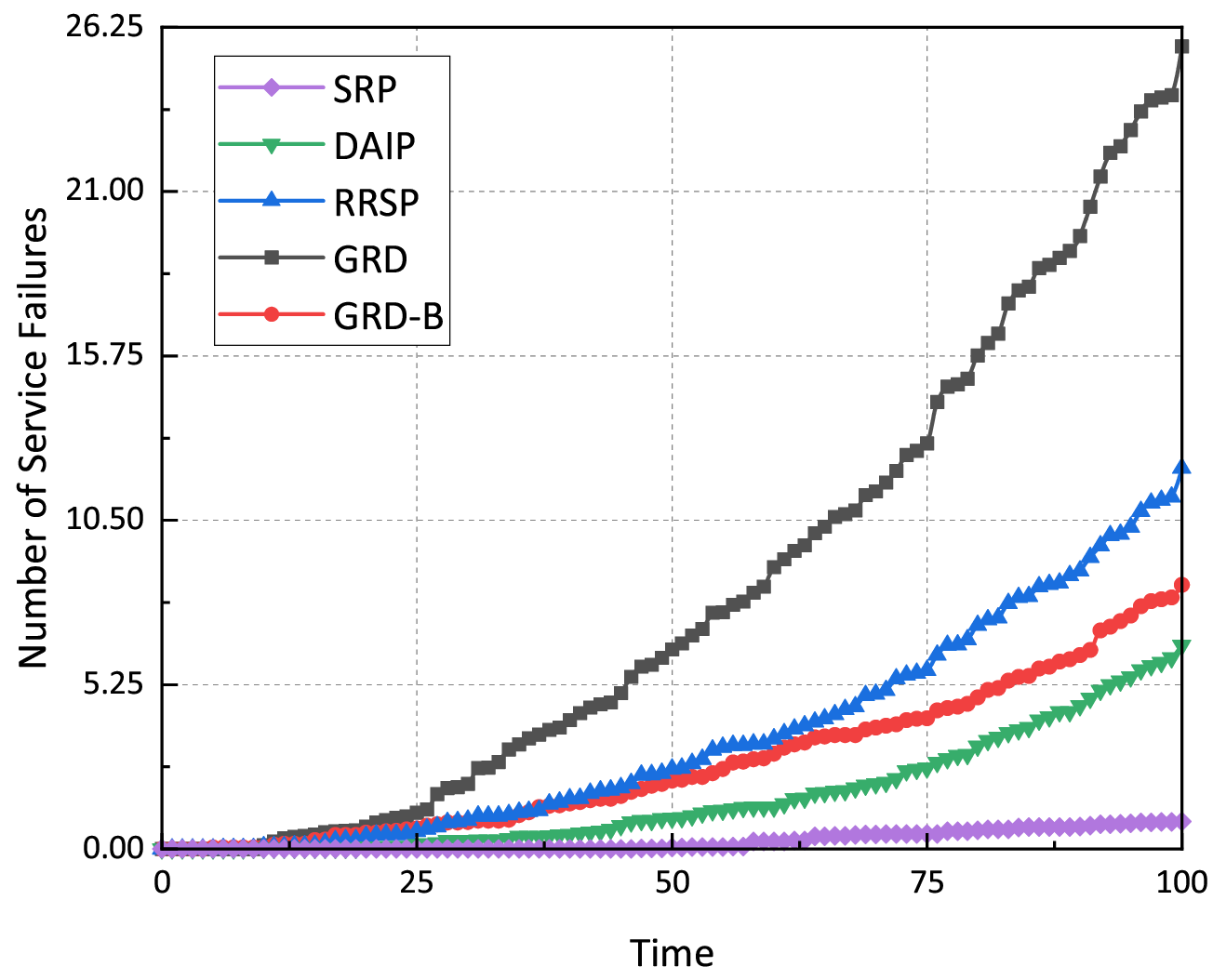}
		\label{fig1b}}
	\caption{Reliability performance with the fully protected path mechanism.}
	\label{fig1}
	\vspace{-1mm}
\end{figure}

\begin{figure}[!t]
	\centering
	\subfloat[Service request distribution for different service reliability.]
	{\includegraphics[width=1.5in]{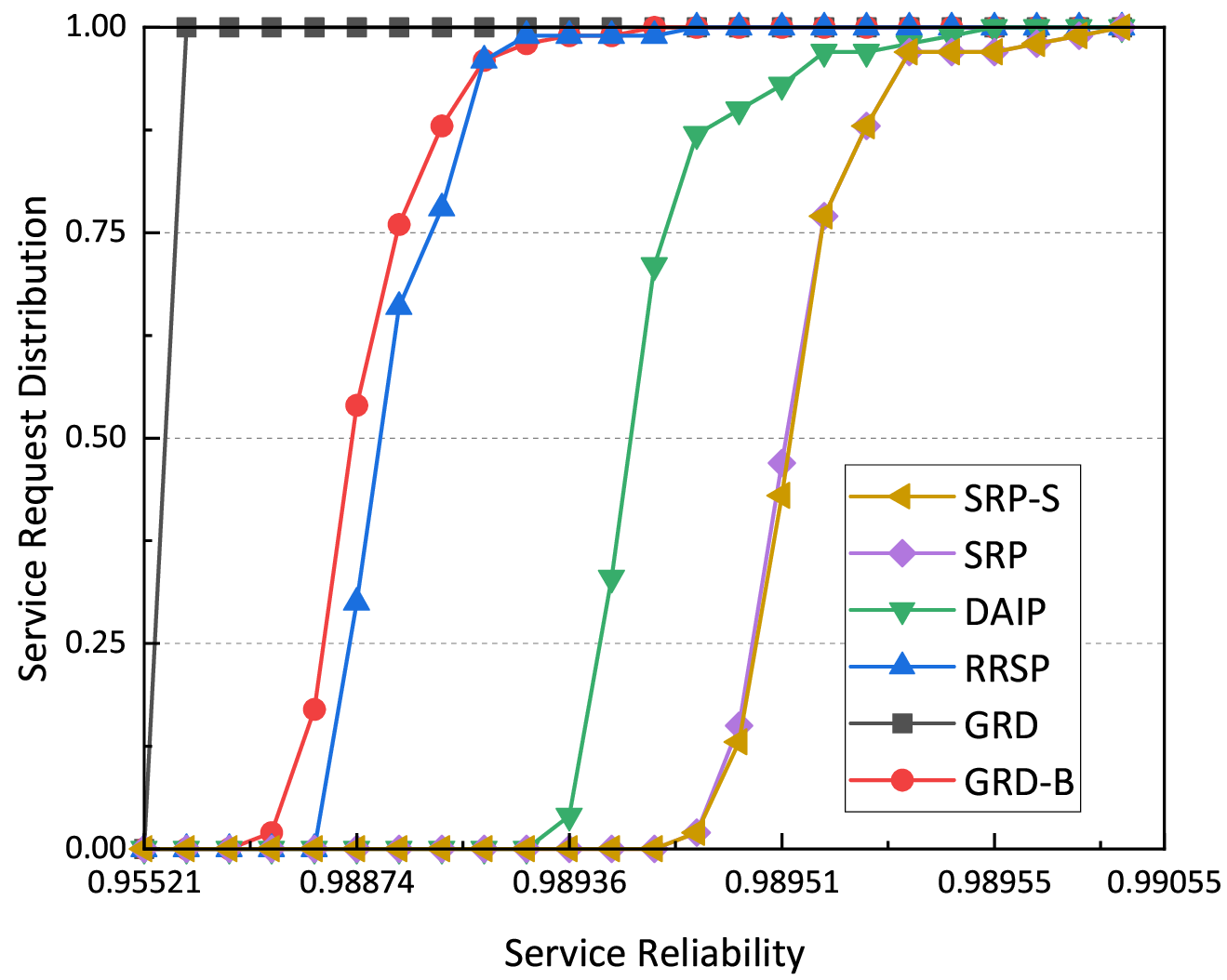}
		\label{fig5a}}
	\hspace{-1.5mm}
	\subfloat[Number of service failures over time.]
	{\includegraphics[width=1.44in]{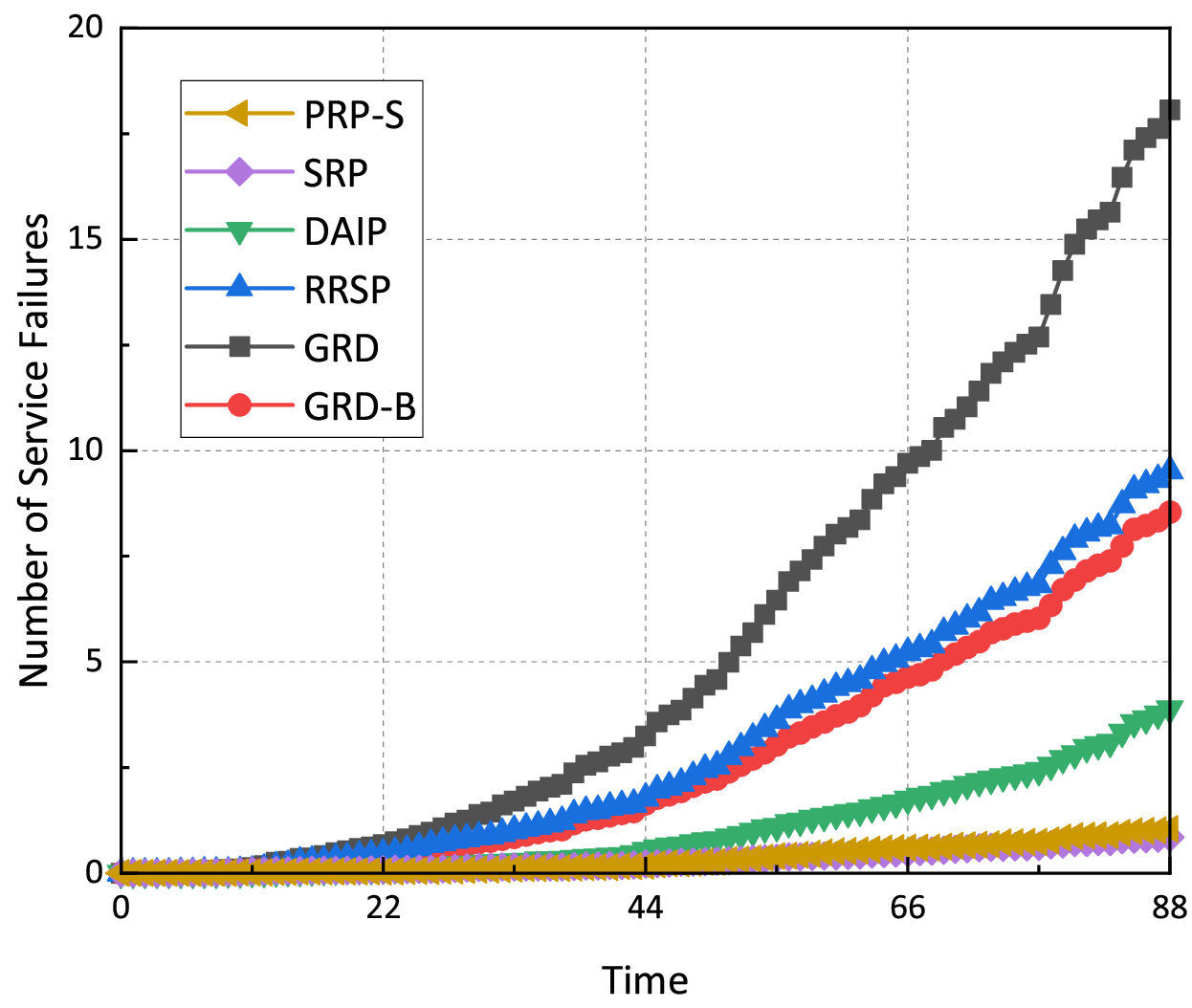}
		\label{fig5b}}
	\vspace{-1mm}
	\caption{Reliability performance with the shared backup path mechanism.}
	\label{fig5}
	
	\vspace{-1mm}
\end{figure}
In this subsection, we verify the effectiveness of the network-aware service reliability model.

Fig. \ref{fig1}(a) and Fig. \ref{fig1}(b) show the network-aware service reliability evaluation and the number of service failures for different placement algorithms with the fully protected path mechanism, respectively. From Fig. \ref{fig1}(a) and Fig. \ref{fig1}(b) we can see that the more service requests with high network-aware service reliability, the fewer number of failures. This result proves that the evaluation result of the proposed reliability model is generally consistent with the evaluation result of the number of failures.

Fig. \ref{fig5} shows the reliability performance with the shared backup path mechanism. As seen in Fig. \ref{fig5}(a) and Fig. \ref{fig5}(b), the proposed network-aware service reliability model can still work to evaluate service reliability with the shared backup path mechanism. This is because when bandwidth resources are not strained, shared backup paths do not need to consider path reliability changes due to backup path contention. However, when bandwidth resources are extremely scarce, the network-aware service reliability will be inaccurate due to backup path contention.
\vspace{-4mm}
\subsection{Validation of the SRP Algorithm}
\vspace{-1mm}
In this subsection, we validate the performance of the SRP algorithm by executing a series of simulations with the fully protected path mechanism. Although we verified the effectiveness of the proposed model in Sec. \ref{sim_model}, a more accurate way to evaluate the performance of the algorithm should still be to evaluate the number of service failures. This is because after different algorithms provide differentiated placement strategies, the resource conditions of the network will gradually differentiate, leading to differences in the solution space for subsequent microservice placement. This makes the reliability of services placed by different algorithms comparable only at the initial moment. Therefore, in order to directly reflect the fault tolerance of different placement algorithms in subsequent simulations, we evaluate the algorithm performance by comparing the number of service failures.

Our first simulation result is shown in Fig. \ref{fig1} (b). The result demonstrates that the SRP algorithm outperforms other algorithms. Taking the number of failures of the worst-performing algorithm as a criterion, the SRP algorithm reduces failures by up to 29\% compared to the latest DAIP algorithm. This is because the proposed algorithm takes into account the network-aware reliability of each placed part when placing microservice instances. In addition, the SRP algorithm always provides a new backup instance for the microservice with the lowest current network-aware reliability when selecting the backup object.
\begin{figure}[!t]
	\centering
	\subfloat[Number of service failures with different edge creation probabilities.]
	{\includegraphics[width=1.4in]{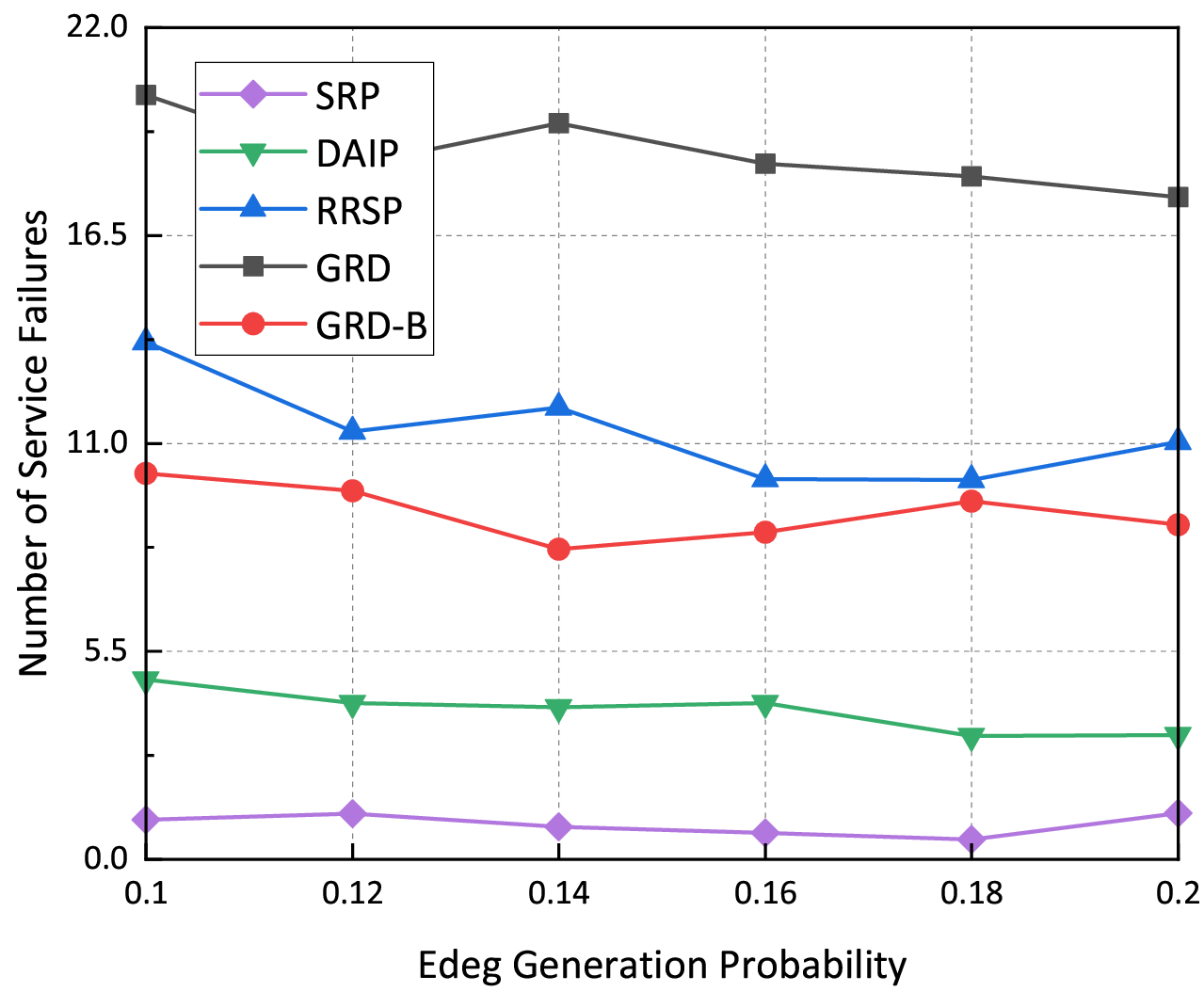}
		\label{fig2a}}
	\hfil
	\subfloat[Number of service failures with a different number of nodes.]
	{\hspace{1mm}\includegraphics[width=1.36in]{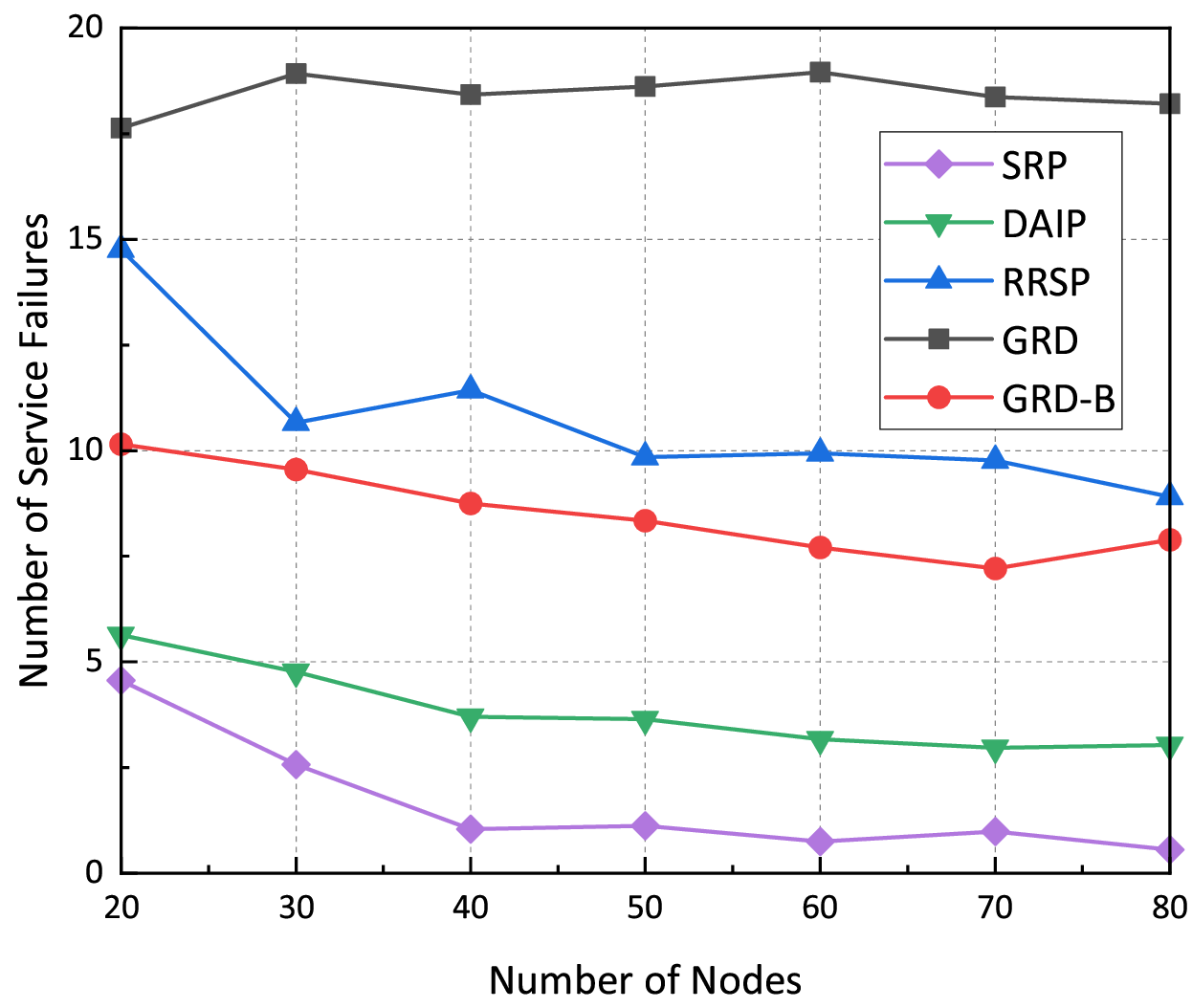}
		\label{fig2b}}
		\vspace{-1mm}
	\caption{Reliability performance with different topologies.}
	\label{fig2}
	\vspace{-1mm}
\end{figure}

We secondly validate the performance of the algorithms with different network topologies. First, we generate six sets of topologies with different edge creation probabilities, and each set has 100 different random topologies. Fig. \ref{fig2}(a) shows the number of failures for services placed by different algorithms with 600 different topologies. From the figure, we can see that the SRP algorithm maintains the lowest number of service failures for all the topologies with different edge creation probabilities. This is due to the fact that the benchmark algorithms all determine the placement of the microservice instances based solely on node reliability, without considering the reduction in reliability caused by network routing and the increase in reliability provided by multipath routing. On the contrary, the SRP algorithm senses network routing by calculating network-aware reliability, thus achieving superior performance. Second, we generated seven sets of topologies with different numbers of nodes, each containing 100 different random topologies. Fig. \ref{fig2}(b) exhibits the number of failures for services placed by different algorithms with 700 different topologies. It can be seen that the number of failures of the different algorithms first decreases and then stabilizes when the number of nodes exceeds 40. This is because the more nodes there are, the more nodes in the network that are in a highly reliable state. Although the benchmark algorithms consider node reliability, they do not consider the reliability of the entire microservice dependency graph. Instead, they select the most reliable node for each instance in isolation, which causes each node to reach a high load state quickly. The SRP algorithm achieves optimal performance because, on the one hand, it calculates the change in node reliability as a node approaches a high load state, and on the other hand, multipath reliability makes distributed placement of instances less costly.

\begin{figure}[!t]
	\centering
	\subfloat[Number of service failures with different CPU requirements.]
	{\includegraphics[width=1.35in]{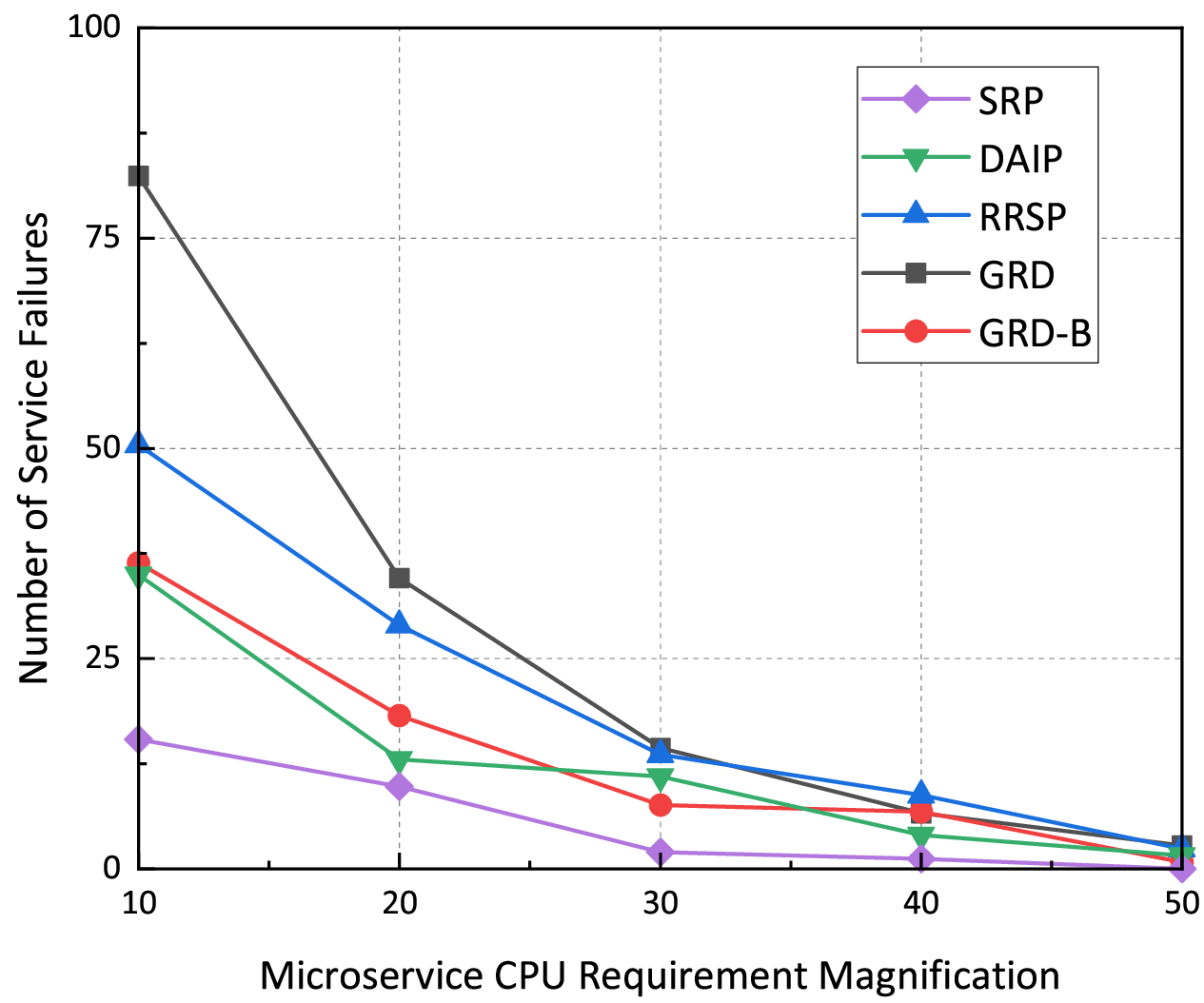}
		\label{fig3a}}
	\hfil
	\subfloat[Number of service failures with different bandwidth requirements.]
	{\includegraphics[width=1.37in]{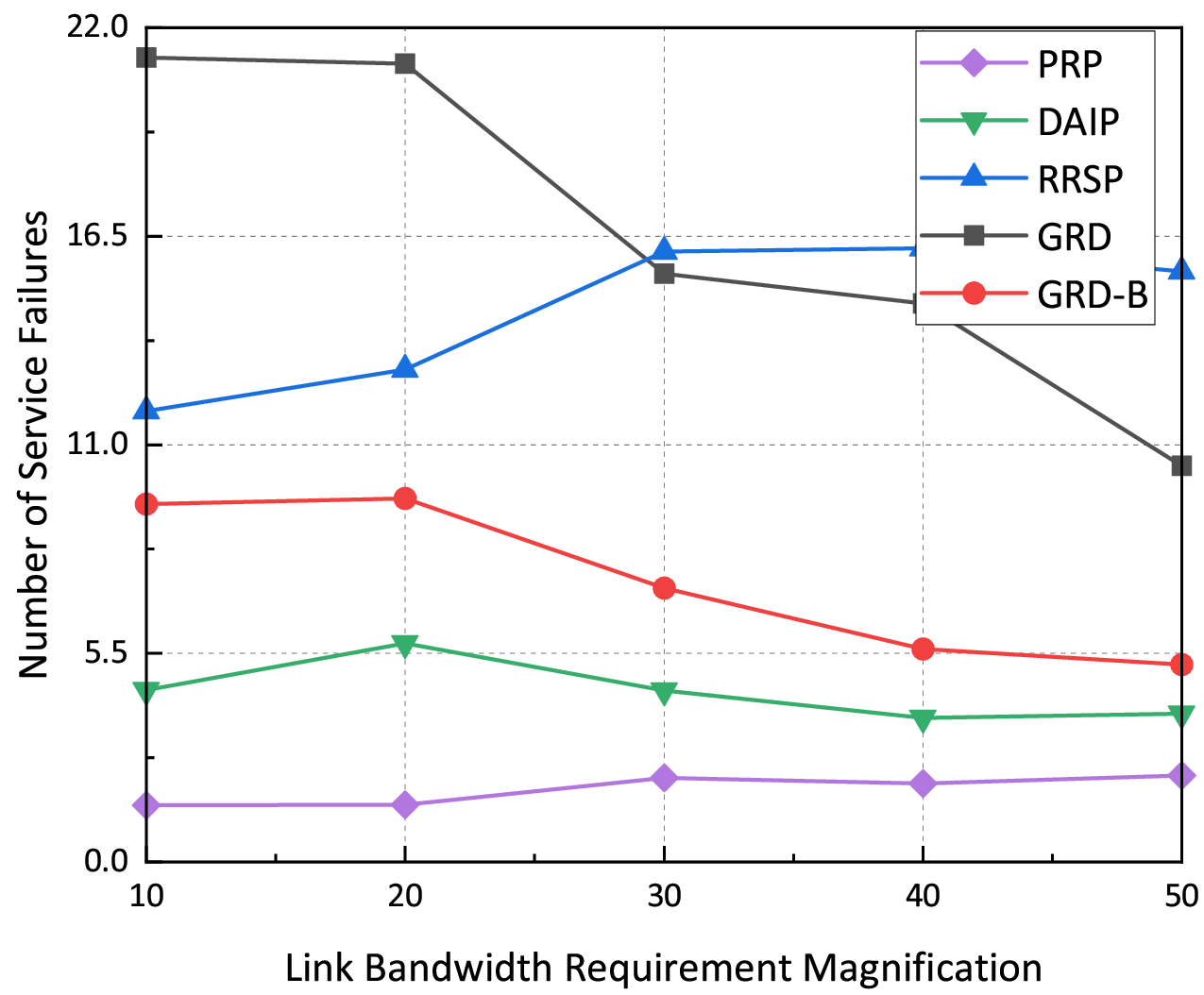}
		\label{fig3b}}
	\vspace{-1mm}
	\caption{Reliability performance with different resource requirements.}
	\label{fig3}
\end{figure}

Thirdly, we verify the performance of the algorithm with different resource requirement conditions. In order to avoid the impact of modifying the topology on the algorithm performance, we do not change the number of nodes in the simulation of Fig. \ref{fig3}(a), but instead scale up the CPU requirements of all the microservices by 10-50 times. The trend of the curve in the figure is decreasing because the number of successfully placed services decreases as the CPU requirement increases, leading to a decrease in the number of service failures. We can see that the SRP algorithm reduces the number of service failures by up to 24\% compared to benchmark algorithms. This is because the benchmark algorithms maintain the strategy of finding the most reliable node when most of the nodes are with high load, whereas the SRP algorithm provides more placement strategies that do not occupy high-load nodes at the cost of occupying the bandwidth of multiple paths. Similarly, we scale up the bandwidth requirement for microservice links in the simulation of Fig. \ref{fig3}(b). We can see that as the bandwidth requirement increases, the number of service failures for most of the benchmark algorithms decreases as fewer services are successfully placed. This is because the inflated bandwidth requirement compresses their solution space. In contrast, the SRP algorithm discovers more solution space through backtracking. Moreover, although the SRP algorithm also places fewer services, it ensures the reliability of successfully placed services through multipath routing and backup. Thus, even though it consumes more bandwidth, the SRP algorithm still achieves superior performance.

\begin{figure}[!t]
	\centering
	\subfloat[Service request distribution for different service reliability.]
	{\includegraphics[width=1.5in]{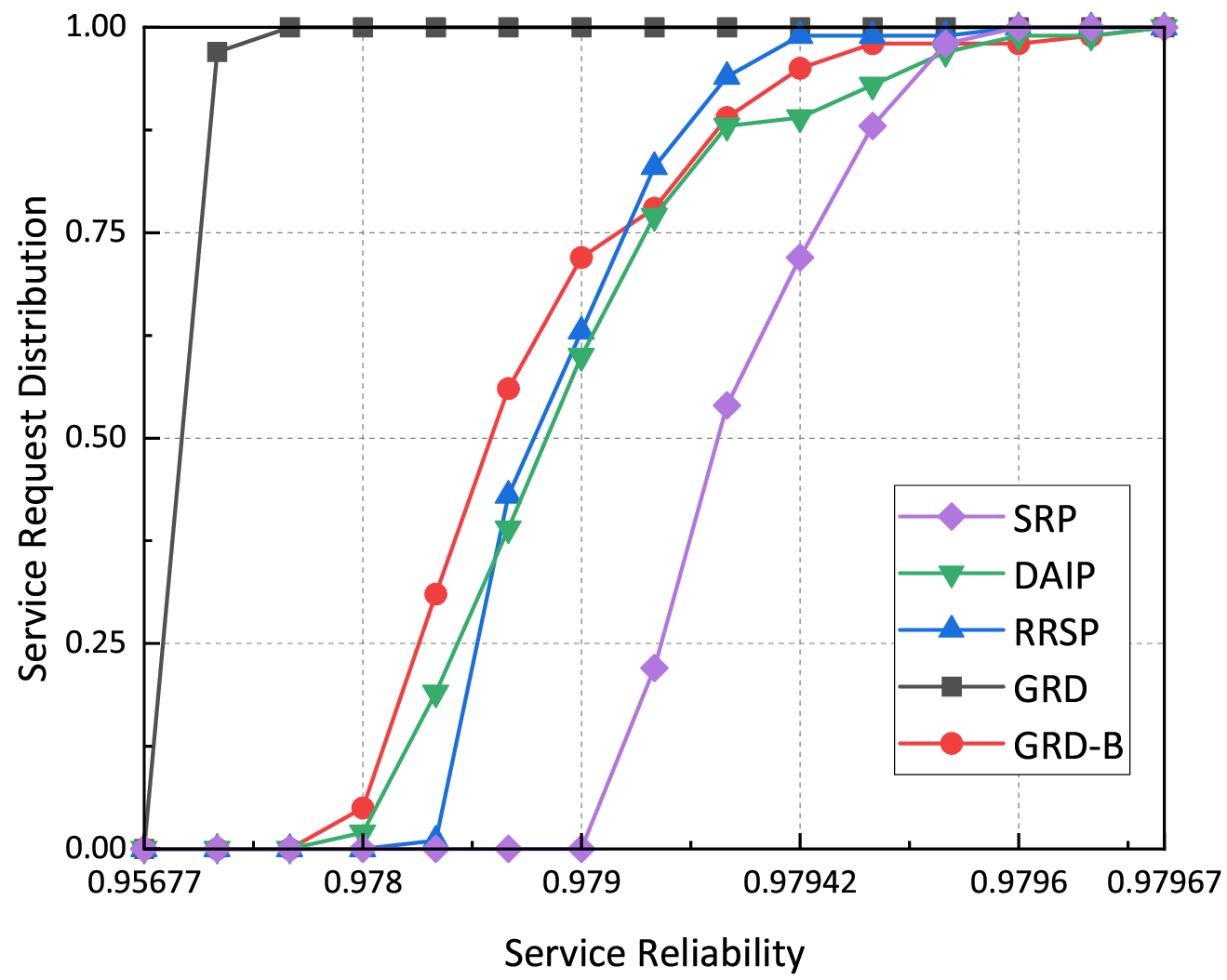}
		\label{fig4a}}
	\subfloat[Number of service failures over time.]
	{\hspace{-1.8mm}\includegraphics[width=1.42in]{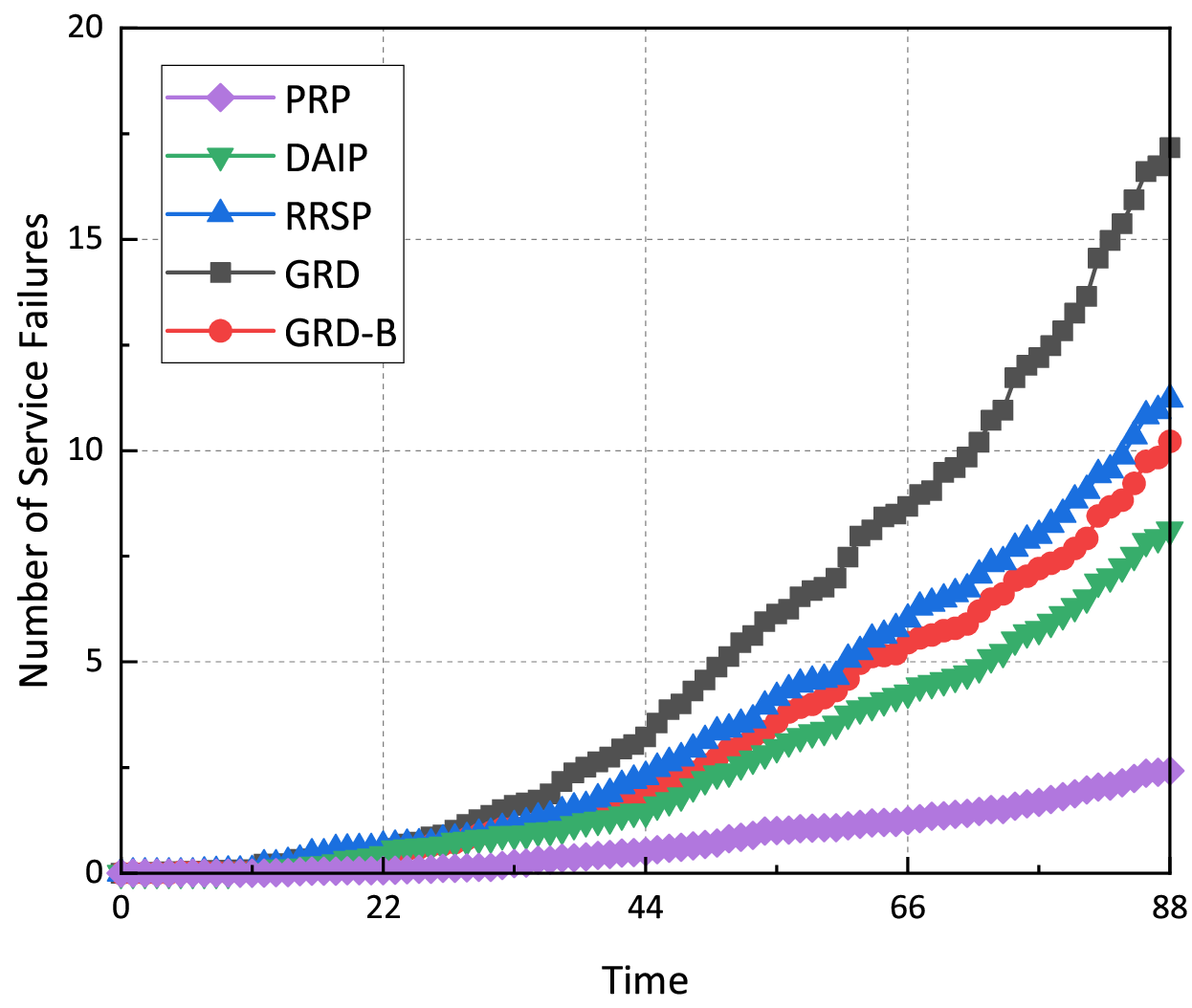}
		\label{fig4b}}
		\vspace{-1mm}
	\caption{Reliability performance with random backups.}
	\label{fig4}
	\vspace{-1mm}
\end{figure}
Finally, we validate the performance of different algorithms for backup object selection. We set the upper limit of backups per service request to a random value between 1 and the number of microservices because when the number of backups decreases, proper backup object selection leads to higher reliability gain. From Fig. \ref{fig4}, we can see that the SRP algorithm can reduce the number of failures by up to 23.8\% despite the reduction in the number of backups compared to the full backup simulation. This relates to the fact that the SRP algorithm considers a backup object selection strategy that prioritizes compensating microservices with the lowest network-aware reliability. 
\vspace{-4mm}
\subsection{Validation of the SRP-S Algorithm}\label{sim_shared}
\vspace{-1mm}
In this subsection, we first verify the significant contribution of the shared backup path mechanism in reducing bandwidth consumption. Then, we verify the fault tolerance performance of the SRP and SRP-S algorithms with the shared backup path mechanism through two simulations.
\begin{figure}[!t]
	\centering
	\includegraphics[width=2.3in]{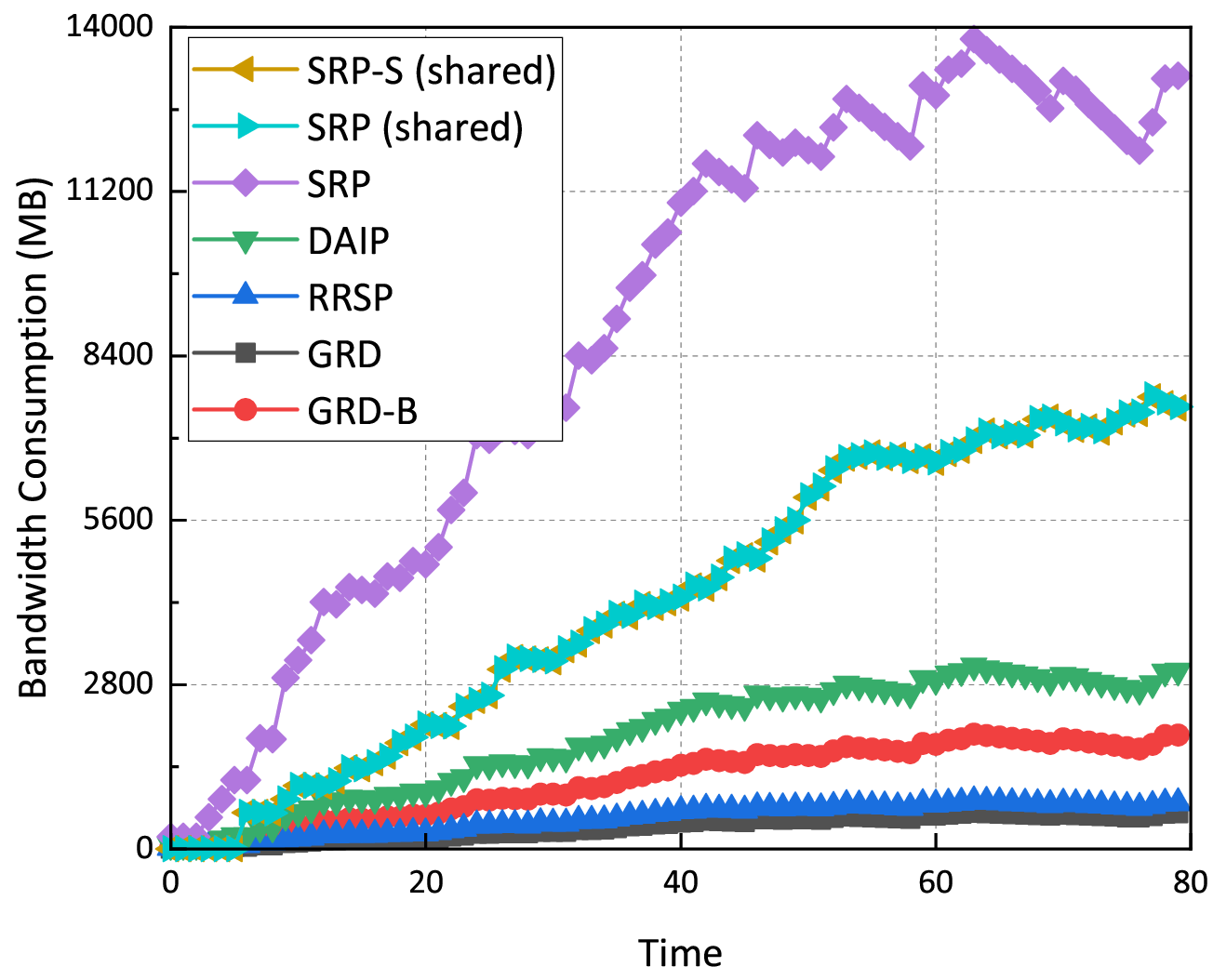}
	\caption{Bandwidth consumption over time.}
	\label{figRc}
	\vspace{-1mm}
\end{figure}

Fig. \ref{figRc} shows the average bandwidth consumption of different algorithms over time, where the algorithms marked with brackets operate with the shared backup path mechanism. As seen in Fig. \ref{figRc}, the SRP-S algorithm is able to reduce the bandwidth consumption by up to 62\% compared to the SRP algorithm with the fully protected path mechanism, which demonstrates its significant potential in reducing bandwidth consumption.

The result of the first simulation for verifying fault tolerance are shown in Fig. \ref{fig5}. From Fig. \ref{fig5}(a) we can see that there is almost no difference between SRP and SRP-S curves. Theoretically, in a single placement with the same conditions, the service reliability of the service request placed by the SRP-S algorithm is not higher than that of the service request placed by the SRP algorithm. This is because the SRP-S algorithm pursues corrected reliability considering the shared backup path contention probability rather than the network-aware service reliability. Fig. \ref{fig5}(b) illustrates the number of service failures of the services placed by each algorithm. In Fig. \ref{fig5}(b), the performance of the benchmark algorithms and the SRP algorithm is similar to the case with the fully protected path mechanism, while the performance of the SRP-S algorithm is similar to that of the SRP algorithm as they are essentially the same when there is not much pressure on bandwidth resources.

\begin{figure}[!t]
	\centering
	\includegraphics[width=2.3in]{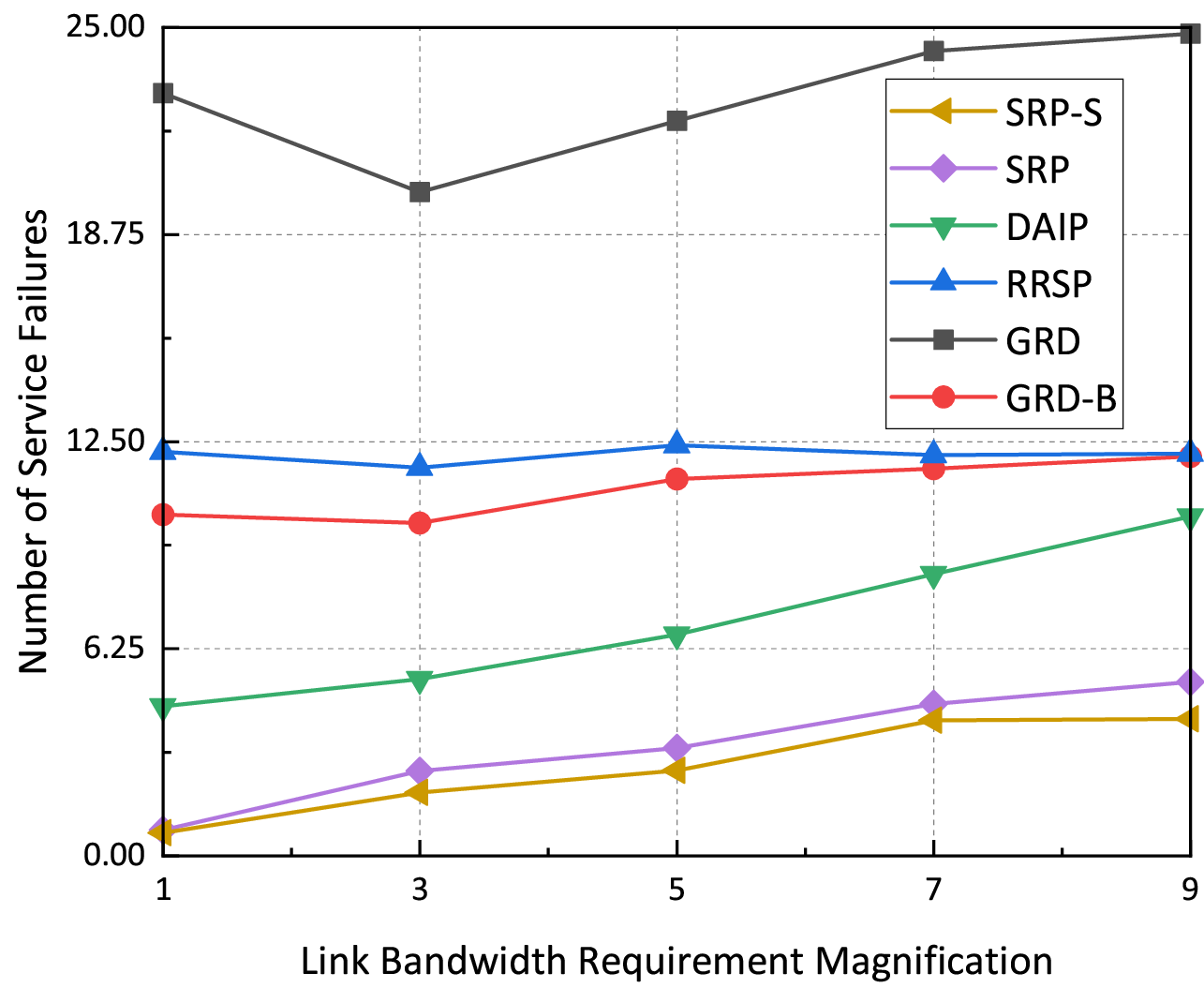}
	\caption{Reliability performance with different bandwidth requirements.}
	\label{fig6}
	\vspace{-1mm}
\end{figure}
To verify the performance of the SRP-S algorithm in more extreme cases, the bandwidth requirements of the microservice links are amplified in the second simulation. From Fig. \ref{fig6} we can see that the SRP algorithm and the SRP-S algorithm produce fewer service failures in different cases and the SRP-S algorithm performs better in the more extreme cases. The reason for the superiority of the SRP algorithm lies in the consideration of network state and backup object selection, whereas the reason for the superiority of the SRP-S algorithm is the consideration of the shared backup path contention caused by simultaneous failure events. Fig. \ref{fig6} shows that the SRP-S algorithm reduces the number of service failures by up to 21\% compared to the SRP algorithm in extreme cases.
	\vspace{-3mm}
\section{Conclusion}
	\vspace{-1mm}
In this paper, we address the intricate challenges of microservice placement with a focus on enhancing the reliability of MSA-based 5G and IoT services. The network-aware service reliability model thoroughly considers the impact of network load and routing on service reliability, offering profound insights into system reliability assessment. Based on the proposed service reliability model, we propose an innovative heuristic SRP algorithm that effectively addresses the microservices placement problem with the fully protected path mechanism. For the purpose of reducing bandwidth consumption, we further propose the SRP-S algorithm by considering the shared backup path contention caused by simultaneous failures, which effectively tackles the microservice placement problem with the shared backup path mechanism. Simulation results validate the proposed service reliability model and show that the SRP algorithm can reduce the number of failures by up to 29\% compared to the benchmark algorithms with the fully protected path mechanism. With the shared backup path mechanism, the SRP-S algorithm can reduce bandwidth consumption by up to 62\% compared to the SRP algorithm with the fully protected path mechanism, and reduce the number of service failures by up to 21\% compared to the SRP algorithm with the shared backup path mechanism.

For future work, we plan to extend our proposed reliability model to more diverse backup mechanisms to reduce bandwidth consumption even further. In addition, it is also one of our goals to adjust the network resource utilization through microservice migration or scaling to improve service reliability in the future.

\vspace{-3mm}
\bibliographystyle{ieeetr}
\bibliography{reference.bib}
\end{document}